\documentclass[11pt, a4paper]{article}
\pdfoutput=1
\usepackage{jcappub}

\usepackage{slashed}
\usepackage{booktabs}
\usepackage{multirow}
\usepackage{amssymb}
\usepackage[export]{adjustbox}
\usepackage{floatrow}
\newfloatcommand{capbtabbox}{table}[][3.5in]
\newfloatcommand{capbfigbox}{figure}[][2.3in]
\usepackage{blindtext}
\usepackage{amsmath}
\usepackage{booktabs}
\usepackage{cancel}
\usepackage{aas_macros}
\usepackage{appendix}
\usepackage{array}
\newcolumntype{x}[1]{>{\centering\arraybackslash}p{#1}}
\usepackage{bm}
\usepackage{color}
\usepackage{graphicx}
\usepackage{cancel}
\usepackage{amsfonts}
\usepackage{amssymb}
\usepackage{amsmath}
\usepackage{calc}
\usepackage{dcolumn}
\usepackage{mathrsfs}
\usepackage{mathtools}
\usepackage{soul}
\usepackage{braket}
\usepackage[normalem]{ulem}

\usepackage[final]{pdfpages}
\usepackage{array}
\usepackage{soul}
\usepackage[normalem]{ulem}
\usepackage{multirow}
\usepackage{pbox}

\pdfoutput=1
\ifx\pdfoutput\undefined
\usepackage[dvips,bookmarks]{hyperref}    % This is for arXiv.org
\else
\usepackage{hyperref}    % This is for pdftex
\fi

     \DeclareMathOperator{\cm}{cm}         \DeclareMathOperator{\few}{few} 
   \newcommand{\cE}{{\cal E}}   \newcommand{\cJ}{{\cal J}} \newcommand{\cL}{{\cal L}} \newcommand{\cM}{{\cal M}}  \newcommand{\cO}{{\cal O}} \newcommand{\cP}{{ \cal P}}   
  
\newcommand{\ie}{{\it i.e.~}}  \newcommand{\eg}{{\it e.g.~}}
    
\newcommand{\pL}{\left(} \newcommand{\pR}{\right)}   \newcommand{\cbL}{\left\{} \newcommand{\cbR}{\right\}}   \newcommand{\ER}{E_R}
\newcommand{\beq}{\begin{equation}} \newcommand{\eeq}{\end{equation}}
\newcommand{\bea}{\begin{eqnarray}} \newcommand{\eea}{\end{eqnarray}}

\newcommand{\tenx}[1]{\times 10^{#1}}
\newcommand{\Eq}[1]{Eq.~(\ref{#1})}  
\newcommand{\Sec}[1]{Sec.~\ref{#1}}  
\newcommand{\Fig}[1]{Fig.~\ref{#1}} \newcommand{\Figs}[2]{Figs.~\ref{#1} and \ref{#2}}
\newcommand{\Tab}[1]{Tab.~\ref{#1}}

\begin{document}
\title{Prospects for Distinguishing Dark Matter Models Using Annual Modulation}
\author[a,b]{Samuel J.~Witte,}
\emailAdd{switte@physics.ucla.edu}
\author[c]{Vera Gluscevic,}
\emailAdd{verag@ias.edu}
\author[d]{and Samuel D.~McDermott}
\emailAdd{samuel.mcdermott@stonybrook.edu}

\affiliation[a]{University of California, Los Angeles, Department of Physics and Astronomy, Los Angeles, CA 90095}

\affiliation[b]{Fermi National Accelerator Laboratory, Center for Particle
Astrophysics, Batavia, IL 60510}

\affiliation[c]{School of Natural Sciences, Institute for Advanced Study, Einstein Drive, Princeton NJ 08540, USA}

\affiliation[d]{C.~N.~Yang Institute for Theoretical Physics, Stony Brook, NY, USA}\subheader{\rm YITP-SB-16-51 \\ FERMILAB-PUB-16-654-A}

\abstract{
It has recently been demonstrated that, in the event of a putative signal in dark matter direct detection experiments, properly identifying the underlying dark matter--nuclei interaction promises to be a challenging task. Given the most optimistic expectations for the number counts of recoil events in the forthcoming Generation 2 experiments, differentiating between interactions that produce distinct features in the recoil energy spectra will only be possible if a strong signal is observed simultaneously on a variety of complementary targets. However, there is a wide range of viable theories that give rise to virtually identical energy spectra, and may only differ by the dependence of the recoil rate on the dark matter velocity. In this work, we investigate how degeneracy between such competing models may be broken by analyzing the time dependence of nuclear recoils, \ie the annual modulation of the rate. For this purpose, we simulate dark matter events for a variety of interactions and experiments, and perform a Bayesian model--selection analysis on all simulated data sets, evaluating the chance of correctly identifying the input model for a given experimental setup. We find that including information on the annual modulation of the rate may significantly enhance the ability of a single target to distinguish dark matter models with nearly degenerate recoil spectra, but only with exposures beyond the expectations of Generation 2 experiments.}

\maketitle

\section{Introduction} \setcounter{page}{2}

A vast array of independent astrophysical and cosmological observations testify to the existence of a non--baryonic form of matter that dominates gravitational potential wells and dictates the dynamics and structure in the universe. However, the particles comprising this dark matter (DM) have so far evaded laboratory probes, despite a direct detection program that has now been mature for several decades. As the next--generation direct detection experiments that incorporate increasingly sensitive detection technologies come online, they will start to probe the final portions of DM parameter space before encountering the so--called `irreducible neutrino background' \cite{Bauer:2013ihz,Aprile:2015uzo,Agnese:2016cpb,Malling:2011va,Newstead:2013pea,Cushman:2013zza,Billard:2013qya,Ruppin:2014bra,Davis:2014ama,Dent:2016iht}. Generation 2 (G2) experiments that are currently, or will soon be, taking data (such as Xenon1T \cite{Aprile:2015uzo}, SuperCDMS SNOLAB \cite{Agnese:2016cpb}, and LZ \cite{Malling:2011va}; see also~\cite{Bauer:2013ihz} for a review) may well be on the cusp of important discoveries, as many interesting theories of DM predict scattering cross sections that live in these portions of parameters space. For example, heavy $SU(2)$--doublet and --triplet fermions, such as the Higgsinos and the wino of supersymmetry, are expected to have cross sections of order $\sigma_{\rm SI} \sim \cO(\few\tenx{-48})\cm^2$ \cite{Hill:2011be,Hill:2013hoa,Hill:2014yxa} (about an order of magnitude below the current limits~\cite{Akerib:2016vxi,Tan:2016zwf}), fixed by their Standard Model gauge quantum numbers alone, while a heavy $SU(2)$--singlet fermion, like the bino, is around an order of magnitude lower depending on its coannihilation partner \cite{Berlin:2015njh}. Models with kinematically suppressed tree--level scattering may also be embedded in more complete dark sectors that have loop--level cross sections in this same range \cite{Ipek:2014gua,McDermott:2014rqa,Appelquist:2015yfa,Appelquist:2015zfa}.

Because so many theories can be accommodated in the parameter space that will be imminently probed by a variety of experiments, it is timely to plan for the science opportunities associated with the first detection of DM particles. Most notably, in case of a confirmed detection, understanding the dark sector dynamics at all energy scales will rely solely on examining low--energy recoils of detector elements and solving the ``inverse problem'' to identify the underlying description of DM--baryon interactions. At the same time, all the information about the dark sector interactions accessible to these measurements is contained within the coefficients of the effective field theory of dark matter direct detection (EFT) \cite{Fitzpatrick:2012ix,Fitzpatrick:2012ib,Anand:2013yka,Bishara:2016hek}. The effective description captures the nontrivial nuclear physics induced by some of the best--motivated UV--complete theories of DM \cite{Gresham:2014vja, Gluscevic:2015sqa,DEramo:2014nmf} through an exhaustible list of nuclear responses that these interactions trigger~\cite{Fitzpatrick:2012ix,Fitzpatrick:2012ib,Anand:2013yka}. It thus provides a systematic framework for classifying and describing a wide variety of DM theories and corresponding phenomenologies observable with direct detection, and we will utilize it in this work. 
 
On the other hand, due to Poisson noise in the number counts of recoil events per unit energy, and similarities in the shape of the nuclear--recoil--energy spectra amongst different interactions, correctly identifying the DM model will present a difficult task in practice, particularly for a single experiment. Recent studies have shown that discriminating between interactions in an agnostic analysis is possible only with strong signals with hundreds of observed recoil events, and only when measurements on targets with sufficiently diverse nuclear physics characteristics are jointly analyzed \cite{Gluscevic:2014vga,Gluscevic:2015sqa} (or, potentially, by jointly analyzing direct and indirect detection data~\cite{Roszkowski:2016bhs}). Thus, using energy spectra to break degeneracies in the DM modeling space crucially relies on complementarity of available target materials \cite{McDermott:2011hx,Peter:2013aha,Gluscevic:2014vga,Catena:2014epa,Catena:2014hla,Dent:2015zpa,Gluscevic:2015sqa,Ruppin:2014bra,Queiroz:2016sxf}, but this still does not guarantee successful model selection if a DM signal is confirmed \cite{Gluscevic:2014vga,Gluscevic:2015sqa,Queiroz:2016sxf}. %Recent work has also shown that dark matter properties may be more accurately identified by jointly analyzing data from direct and indirect dark matter searches~\cite{Roszkowski:2016bhs}. 

Almost since the dawn of direct detection related DM studies, the motion of Earth relative to DM bound in the galactic halo has been predicted to provide a distinctive signature of DM through annual modulation of the nuclear recoil rate~\cite{Freese:1987wu, Freese:2012xd,Lee:2013xxa,Britto:2014wga,DelNobile:2015nua,Kouvaris:2015xga}. While the annual modulation signal is typically assumed to take an approximately experiment--independent form, recent work has pointed out that non--standard interaction cross sections could produce a modulation signal that is unique to each target element \cite{DelNobile:2015tza,DelNobile:2015rmp}. More generally, a non--trivial velocity dependence in the cross section effectively changes the phase space integral that governs the total event rate of recoil events in a given experiment, producing a non--standard modulation signal. Thus, it may be expected that interactions differing solely by the DM velocity dependence of their corresponding cross sections may give rise to different phase and/or amplitude of the annual modulation signal. 

Motivated by this argument, we propose here that an analysis of the time dependence of scattering events can help discriminate between interaction models whose recoil energy spectra are otherwise degenerate on a single target material. Using the method of \cite{Gluscevic:2015sqa}, we create a suite of simulations under a variety of scattering theories, and apply a Bayesian model selection analysis on the simulated data to evaluate the chance for correctly identifying the underlying model. We statistically evaluate the enhancement in prospects for accurate model selection when the annual modulation signal is analyzed in combination with recoil--energy measurements in the future--generation direct detection experiments. 

The rest of the paper is organized as follows. In \Sec{sec:dd} we review the calculation of the direct detection scattering rate and discuss how direct detection observables (including the annual modulation) differ depending on the momentum and velocity dependence of the interaction. \Sec{sec:procedure} summarizes the models and experiments considered in this work, and describes our simulations and analysis method. We present the results of this analysis in \Sec{sec:results}, and conclude in \Sec{sec:conclusion}.

\section{Scattering in Direct Detection Experiments}\label{sec:dd}

In this Section, we provide an overview of the calculation of the nuclear recoil rate in direct detection experiments, specifically focusing here on elastic scattering. We then summarize potential novel momentum and velocity dependent scattering cross sections, categorized by the EFT approach, and discuss possible phenomenologies observable in direct detection experiments. Finally, we illustrate how different velocity dependencies may give rise to distinct time dependent signatures in the recoil data.

\subsection{Rate Calculation}\label{sec:rate}

The key measurement of most direct detection experiments is the nuclear recoil energy spectrum --- the number count of nuclear recoil events per recoil energy $E_R$, per unit time $t$, per unit target mass, which reads
\begin{equation}
\frac{dR}{dE_R dt}(E_R,t) =  \frac{\rho_\chi}{m_T m_\chi} \int\limits_{v_{\mathrm{min}}}^{v_{\mathrm{esc, lab}}}  v f(\mathrm{\mbox{\bf{v}}},t) \frac{d\sigma_T}{dE_R} (E_R,v) d^3v \, .
\label{eq:dRdEr_general}
\end{equation}
Here, $\rho_\chi$ is the local DM density; $m_\chi$ is the DM particle mass; $m_T$ is the mass of the target nucleus $T$; $\mathrm{\mbox{\bf{v}}}$ is DM velocity vector of magnitude $v$ (in the lab frame); $f(\mathrm{\mbox{\bf{v}}}, t)$ is the observed DM velocity distribution; $d\sigma_T/dE_R=m_T \sigma_T /2\mu_T^2 v^2$ is the differential cross section for DM scattering off a nucleus $T$; and $\mu_T\equiv\frac{m_Tm_\chi}{m_T+m_\chi}$ is the reduced mass of the DM particle and the target nucleus. Integration limits are the minimum velocity a DM particle requires in order to produce a nuclear recoil of energy $E_R$, given by $v_\mathrm{min} = \sqrt{m_T E_R/2\mu_T^2}$, and the Galactic escape velocity in the lab frame, $v_{\mathrm{esc, lab}}$. Here, we define the overall normalization of $\sigma_T$ as $\sigma_p$, and refer the interested reader to \cite{Gluscevic:2015sqa} for the interaction--dependent definitions (note that we also list these definitions in the final column of Table~\ref{tab:operators}). We note that $\sigma_p$ is one of the key free parameters of each scattering interaction.

The differential rate in \Eq{eq:dRdEr_general} is determined by the experimental setup, the DM astrophysical and particle properties, the nuclear properties of the target material, and the DM--nucleus interaction. For the purposes of this study, we set the astrophysical parameters to the following values \cite{Bovy:2013raa,Piffl:2013mla}: $\rho_\chi=0.3$ GeV/cm$^3$; $v_{\mathrm{esc}} = 533$ km/sec (in the Galactic frame), and assume that $f(\mathrm{\mbox{\bf{v}}})$ is a Maxwellian distribution in the Galactic frame, with a rms speed of $155$ km/sec and a mean speed equal to the Sun's rotational velocity around the Galactic center, $v_\textrm{lag}=220$ km/sec.
The underlying particle physics interaction determines the calculation of the recoil rate through the differential scattering cross section ${d\sigma_T}/{dE_R}$ \cite{Gluscevic:2015sqa,Gresham:2014vja}. Different interactions display different functional dependences on $E_R$ and $v$, as discussed in detail in Refs.~\cite{Gluscevic:2015sqa,Gresham:2014vja} and summarized below in \Sec{subsec:momentum_velocity}.

The total rate $R$ of nuclear recoil events (per unit time and unit mass) is given by the integral of the differential rate within the nuclear--recoil energy window $\cE$ of a given experiment, $R(t)=\int_\cE \frac{dR}{dE_R dt} dE_R$. For simplicity, we assume unit efficiency of detection within the analysis window, and rescale individual experimental exposures to take this assumption into account when choosing experimental parameters to represent the capabilities of G2 experiments. In turn, the total expected number of events $\langle N\rangle$ for a fiducial target mass $\cM_\textrm{fid}$, in experiment that started observation at a time $t_1$ and ended at a time $t_2$, is given by
\beq \label{eq:totevents}
\langle N_\mathrm{tot}\rangle =  \cM_\textrm{fid} \int\limits_{t_1}^{t_2} \int\limits_\cE  \frac{dR}{dE_R dt}(E_R,t)\,dE_R \,dt.
\eeq

\subsection{Momentum and Velocity Dependence}
\label{subsec:momentum_velocity}

Traditional focus on the two standard scattering cases, spin--independent (SI) and spin--dependent (SD) scattering (the former involves coherent contributions from the entire nucleus, resulting in a cross section that scales quadratically with nucleon number, while the latter scales with the total nuclear spin), obscures the richness of phenomenologies which can arise when these two standard interactions are suppressed \cite{Fitzpatrick:2012ix,Gresham:2014vja}. Here we summarize the EFT that catalogues all possible energy and velocity dependencies of the cross section, and thus delineates the modeling space for interactions probed by these experiments, in most general terms. In the \Sec{sec:models}, we highlight several well--motivated examples of scattering models which we use in this work to examine the extent to which including time information can help identification of the underlying DM model.

The EFT of DM direct detection \cite{Fitzpatrick:2012ix, Anand:2013yka} relies on an expansion in two small kinematic variables: $|\vec q|/m_n$ and $|\vec v_\perp|$; $\vec q$ is the change in momentum of the DM particle during the scattering, related to the recoil energy as $\vec{q}^{\, 2} =2m_TE_R$, $|\vec v_\perp|$ is the component of the relative velocity of the initial--state particles that is orthogonal to the momentum transfer, and $m_n$ is the mass of the nucleon. For an incoming (outgoing) DM three--momentum $\vec p$ $(\vec p\,')$, incoming (outgoing) nuclear three--momentum $\vec k$ $(\vec k')$, and a DM--nucleon reduced mass $\mu_{\chi n}$, these factors read $\vec q=\vec p\,'-\vec p=\vec k-\vec k'$, and $\vec v_\perp=\frac{\vec p}{m_\chi}-\frac{\vec k}{m_n}+\frac{\vec q}{2\mu_{\chi n}}$.

These expansion parameters are of the same order of magnitude, but it is important to note that they manifest differently in the observables of the scattering events (see \eg \cite{Gluscevic:2015sqa} for a comprehensive discussion). In particular, terms that enter at higher order in $|\vec q|/m_n$ deliver a vanishing event rate at both small and large momentum transfer (or, equivalently, recoil energy), with a maximum rate at some intermediate recoil energy, producing a ``turnover'' feature in the spectrum. Light mediator models can alternatively contain factors of $m_n/|\vec q|$, producing a steep enhancement of the recoil events at low values of $\ER$. On the other hand, higher--order terms in $| \vec v_\perp|$ produce event rates that monotonically decrease with recoil energy, similar to the case of the standard SI and SD interactions (see \Fig{fig:diff_rate_comp} for illustration).

As was demonstrated by Ref.~\cite{Gluscevic:2015sqa}, interactions that feature different momentum dependence can be differentiated from each other using a single nuclear target, provided a sufficiently large number of events are observed; however, the latter class of models --- those that differ only by the power of velocity dependence --- are far more difficult to disentangle, leaving substantial degeneracy between well--motivated models. In the following, we develop an intuition for how this degeneracy might be overcome, using annual modulation and time dependence of the scattering rate.

\subsection{Time Dependence}
\label{subsec:time}

In \Eq{eq:dRdEr_general}, the differential rate of nuclear recoils is explicitly denoted as depending on time, which arises as a consequence of the Earth's harmonic motion around the Sun. This motion causes the total DM particle flux observable by direct detection experiments to modulate at the few percent level. The expected phase and amplitude of the modulation depend on the astrophysical and particle properties of DM (see \eg \cite{Green:2000ga,Gelmini:2000dm,DelNobile:2015nua}); they are additionally modified by the effect of gravitational focusing of DM by the Sun, which produces a characteristic energy dependence in the phase of the modulation~\cite{Danby01021957,Griest:1987vc,Sikivie:2002bj,Alenazi:2006wu,Lee:2013wza,DelNobile:2015nua}. 

\begin{figure*}
\centering
\includegraphics[width=0.49\textwidth, trim=0.cm 0.0cm 0.cm 0.0cm,clip=true]{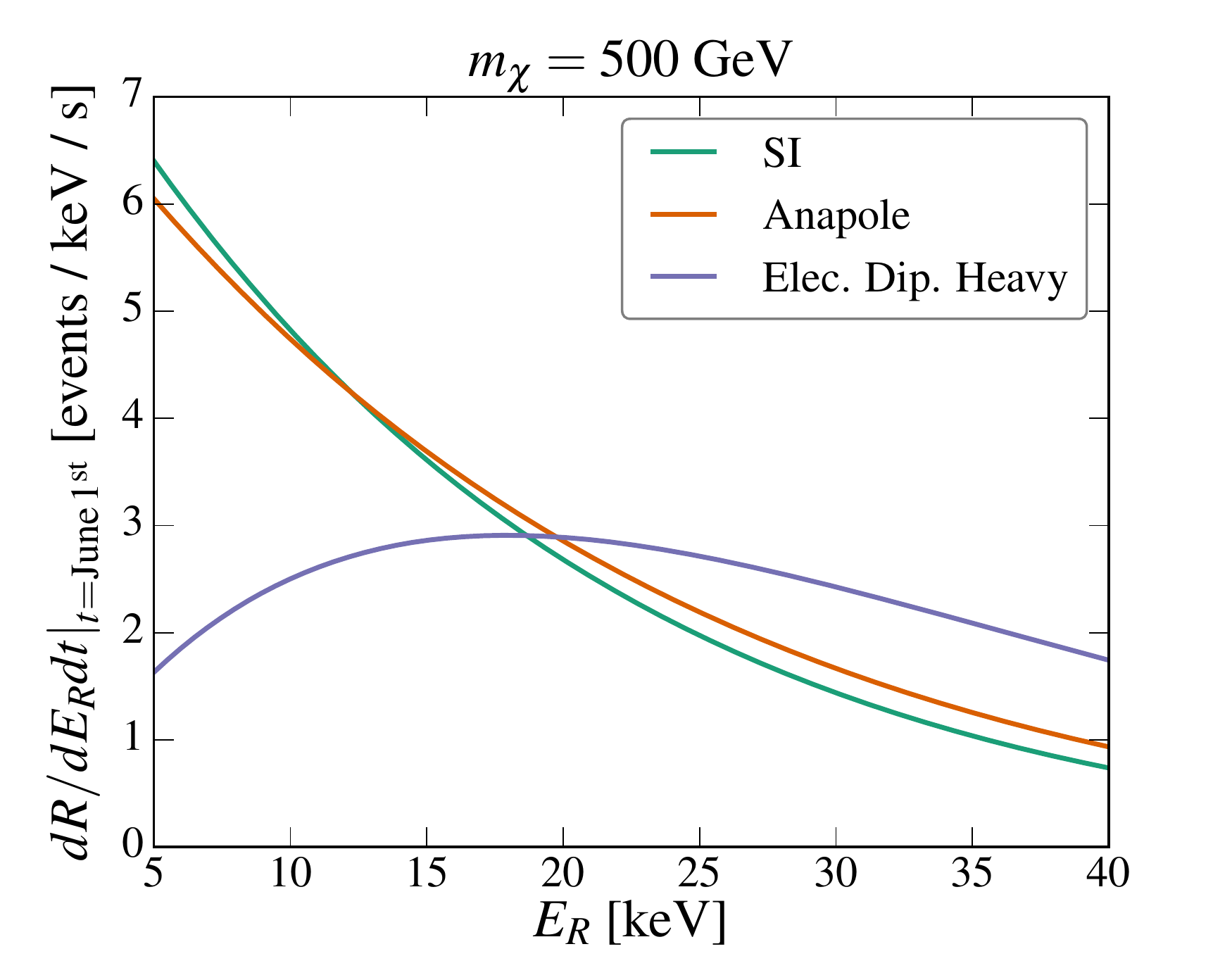}
\includegraphics[width=0.49\textwidth, trim=0.cm 0.0cm 0.cm 0.0cm,clip=true]{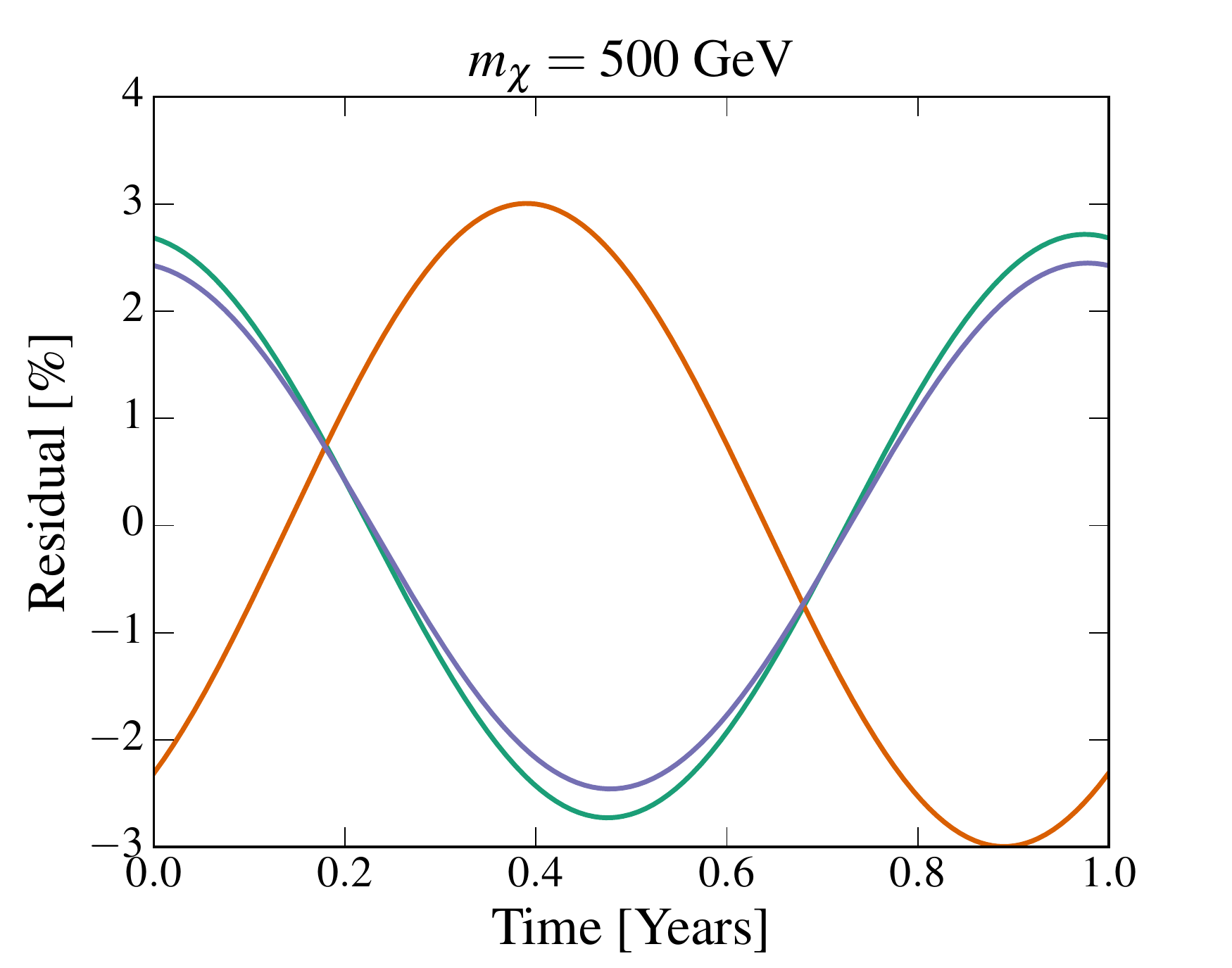}
\includegraphics[width=0.49\textwidth, trim=0.cm 0.0cm 0.cm 0.0cm,clip=true]{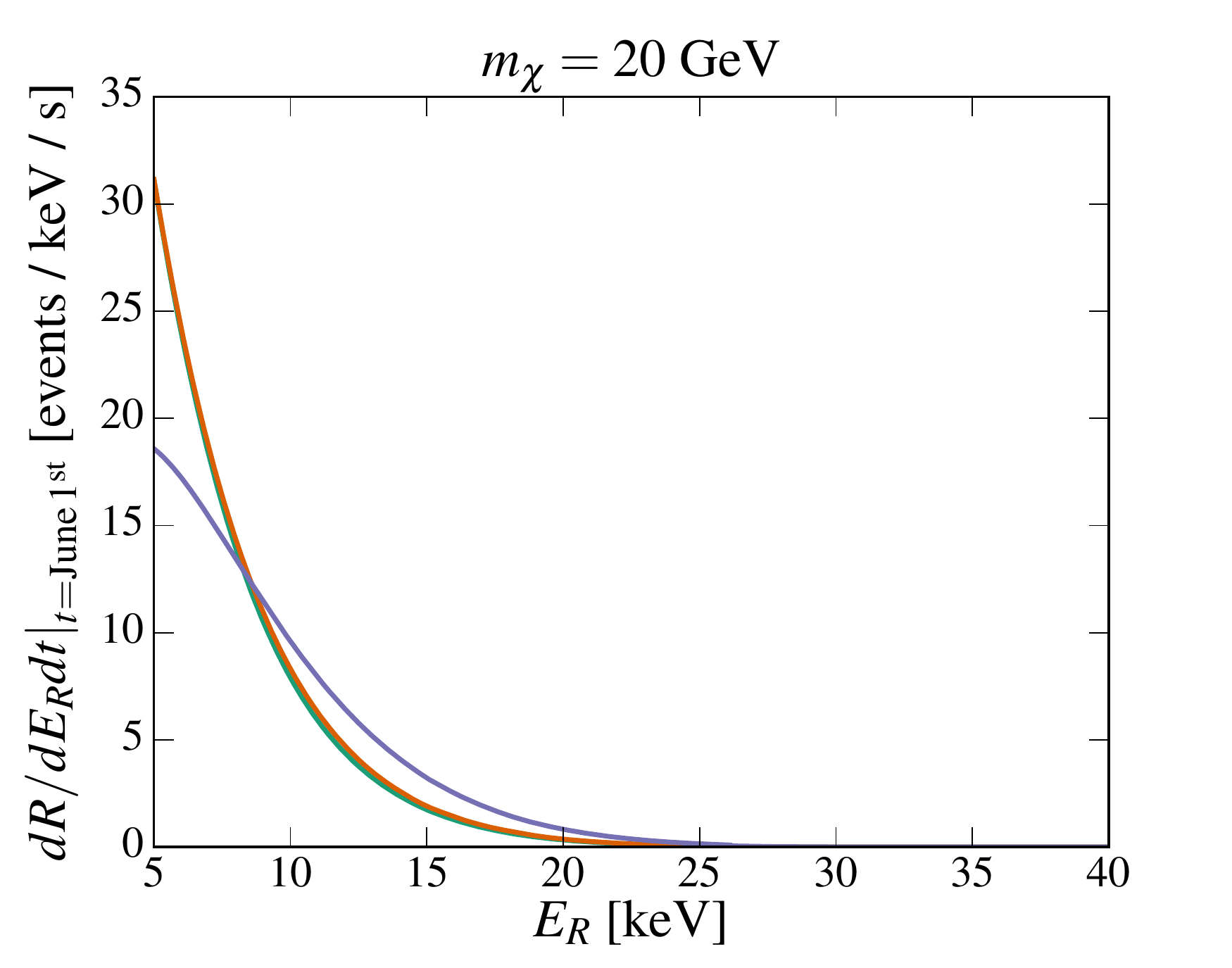}
\includegraphics[width=0.49\textwidth, trim=0.cm 0.0cm 0.cm 0.0cm,clip=true]{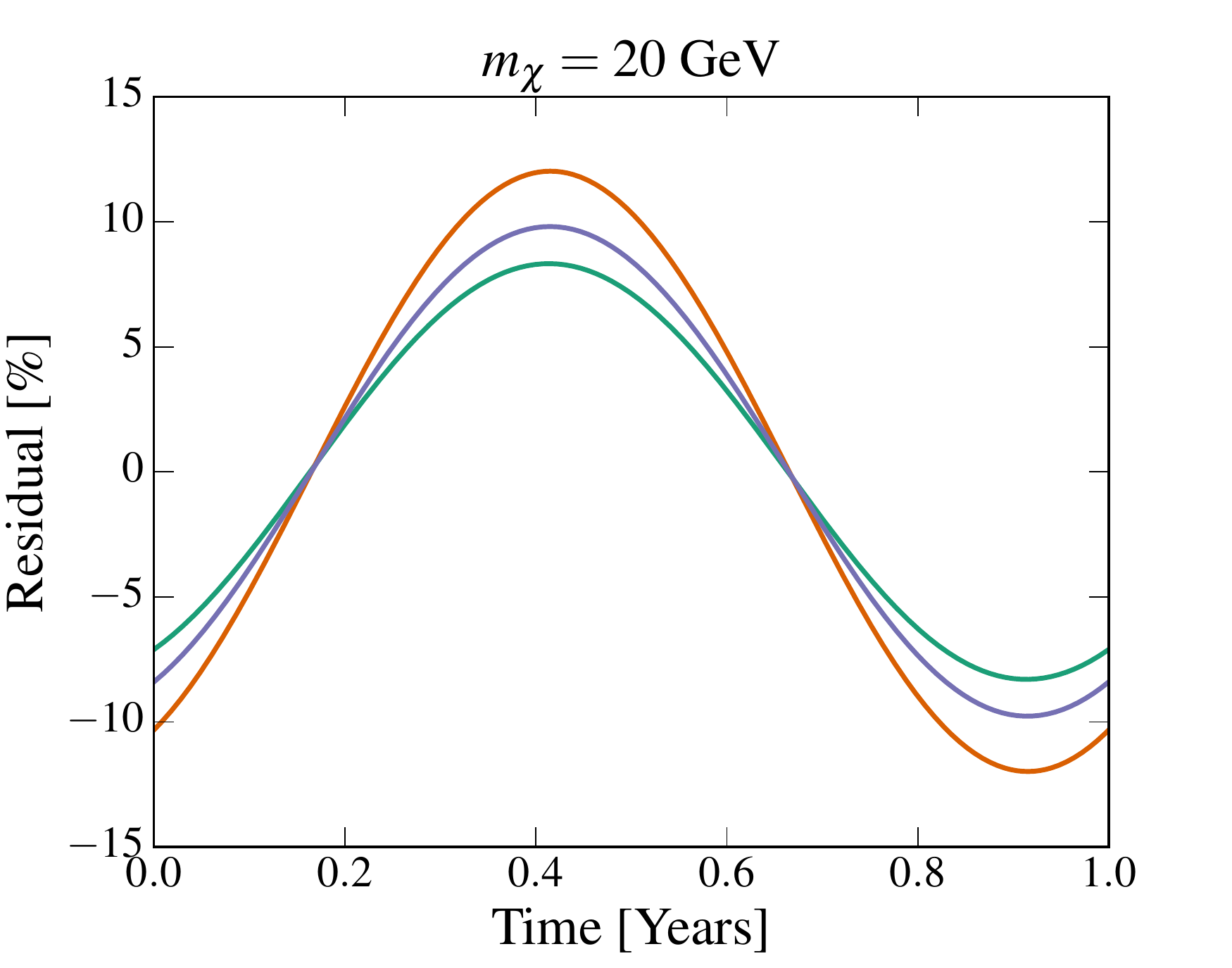}
\caption{\label{fig:diff_rate_comp}
Comparison of the nuclear recoil energy spectra (left column) and annual modulation signals (right column) between the SI, anapole, and heavy--mediator electric dipole interaction models on a xenon target, where the cross sections have been normalized to the current LUX 90\% confidence level exclusion limit~\cite{Akerib:2016vxi}. Top row corresponds to a $500$ GeV, and the bottom row to a $20$ GeV DM particle. \emph{Left:} Differential event rate (evaluated for June $1^{st}$) as a function of the nuclear recoil energy. \emph{Right:} Residual event rate (fractional deviation in the total event rate) as a function of time.}
\end{figure*}

We illustrate differences in the recoil energy spectra and in the annual modulation signal in the context of interactions with a differing functional dependence on the momentum transfer and DM velocity in \Fig{fig:diff_rate_comp}, for several DM--nuclei interaction models. Specifically, we compare the standard SI, anapole, and heavy--mediator electric dipole (ED--heavy) interactions (see \Sec{sec:models} for a more detailed definition of the models). The top row of panels corresponds to a 500 GeV, and the bottom row to a 20 GeV DM particle.  Note that the energy spectra for the SI and ED--heavy interactions are distinct in a way that the SI and anapole interactions are not; thus, discerning the SI and anapole hypotheses using the energy spectrum alone is quite challenging, given even the most optimistic expectations for the Poisson noise \cite{Gluscevic:2015sqa}. 

However, the annual modulation of the standard SI and anapole interactions can be very different, owing to a non--trivial ($\sim | \vec v_\perp|^2$) velocity dependence of the anapole cross section. This non--trivial velocity dependence in turn alters the velocity integral in \Eq{eq:dRdEr_general}, and consequently leads to a different time dependence of the total event rate in the two interaction models (see also Figs.~1--3 of \cite{DelNobile:2015rmp} for an overview of time--dependent behavior of various $v$--dependent cross sections). For large enough DM mass, this effect can produce a nearly {\it opposite modulation phase} between the standard SI scenario and the anapole case. Furthermore, differential cross sections which contain multiple non--negligible terms with different velocity dependences can produce annual modulation signals entirely unique to a given target element~\cite{DelNobile:2015tza,DelNobile:2015rmp}. The time variation of the rate, and thus the differences between the annual modulation produced by different interactions, is typically expected to be small --- on the order of a few percent. Nonetheless, we will show in the following that this small difference can be used to supplement the information contained in the energy spectrum and substantially aid the process of model selection using a single target element. 

\section{Distinguishing Scattering Models}\label{sec:procedure}

Our approach, outlined below, follows that of Ref.~\cite{Gluscevic:2015sqa}. To address our main question, we begin by selecting several well--motivated scattering models featuring a similar dependence on the momentum transfer, but a different dependence on the DM velocity (\ie models with nearly degenerate recoil spectra but qualitatively different annual modulation); we summarize our choice of models in \Sec{sec:models}. Then, we simulate nuclear--recoil events under these models, for three different DM masses. For our simulations, described in detail in \Sec{sec:sims}, we use cross sections that are at the current exclusion limit for a given interaction at hand. In order to capture in the impact of Poisson noise on future data analyses, we create a suite of simulations for each choice of model, mass, cross section, and target element. We then perform Bayesian model selection (described in detail in \Sec{sec:stats}) between two competing models (hypotheses) --- the one used to create the simulation (``true underlying model'') and the competing model (\ie a ``wrong'' model) that has a nearly degenerate recoil spectrum but different time dependence.  We repeat this procedure on each simulation in a given suite, to evaluate chances that future data confidently selects the underlying model. Model selection is repeated two times for each simulation --- once including and once neglecting the time dependence of the rate (\ie the annual modulation) in the likelihood function. Comparison of the two corresponding results enables us to quantitatively assess the impact that the inclusion of time information may have on prospects for identifying the true model.

\subsection{Summary of Models}\label{sec:models}

Here, we illustrate a generic scenario which gives rise to the DM--nuclei interactions we consider in this paper. We emphasize that this scenario by no means represents and exhaustive list of possible, or even well--motivated, models, but is rather just an illustrative example for studying the operators we are interested in (for a more comprehensive discussion we refer the interested reader to~\cite{Gresham:2014vja,Gluscevic:2015sqa}). 

We thus focus on a generic extension of the Standard Model, represented by a hidden $U(1)'$ that has several charged fermions $\psi_i$ and a heavy gauge boson $A'_\mu$ with mass $M$ that kinetically mixes with the Standard Model photon. At high energies, the Lagrangian contains
\beq \label{eq:UV-model}
\cL \supset -  m_i \bar{\psi}_i \psi^i + i \bar \psi_i \slashed D_{ij} \psi^j  - \frac12 M^2 A'_\mu A'^\mu  - \frac14 F'_{\mu \nu} F'^{\mu \nu} - \frac\epsilon2 F'_{\mu \nu} F^{\mu \nu} \, ,
\eeq
where $F_{\mu \nu}$ and $F'_{\mu \nu}$ are the field strength tensor of the photon and the heavy gauge boson, respectively (\ie $F_{\mu \nu} \equiv \partial_\mu A_\nu - \partial_\mu A_\nu$). At low energies, the $A'_\mu$ and most $\psi$ particles are integrated out. We assume a mass hierarchy that results in an electrically neutral fermion $\chi$ as the lightest degree of freedom in the dark sector, thereby providing a DM candidate. Because of the kinetic mixing, the state $\chi$ couples to the Standard Model nucleon current \cite{Gresham:2014vja},
\beq \label{eq:current}
\cJ_\mu = \partial^\alpha F_{\alpha \mu} = e \sum \bar n \pL Q_{n} \frac{K_\mu}{2m_{n}} -\widetilde \mu_{n} \frac{i \sigma_{\mu \nu}q^\nu}{2m_n} \pR n,
\eeq 
where the sum runs over individual nucleons, $Q_n=1(0)$ are proton (neutron) charges in units of the electron charge $e$, $K_\mu/2 = (k_\mu + k'_\mu)/2$ is the average nucleon momentum, and $\tilde{\mu}_n = {\text{magnetic moment} \over \text{nuclear magneton}}$ is the dimensionless magnetic moment of the nucleon.

The details of the masses and charges of the dark fermions $\psi_i$ that constitute or couple to the DM $\chi$ will determine interaction that is measured in an experiment. We will use $\cO_\chi^\mu$ to denote the Lorentz--vector fermion bilinear that couples to the current in \Eq{eq:current}. Because we assume $\chi$ is electromagnetically neutral, the possible $\cO_\chi^\mu$ are \cite{Gresham:2014vja, Gluscevic:2015sqa}
\begin{eqnarray} \label{eq:photon-DM-ops}
\cO_{\rm \chi, Anapole}^\mu & = & g^{\rm Anapole}\bar \chi \gamma^\mu \gamma_5 \chi, \\
\cO_{\rm \chi, MD}^\mu & = & \frac{g^{\rm MD}}{\Lambda}\bar \chi i \sigma^{\mu \nu} q_\nu \chi ,\\
\cO_{\rm \chi, ED}^\mu & = & \frac{g^{\rm ED}}{\Lambda} \bar \chi i \sigma^{\mu \nu} \gamma_5 q_\nu \chi.
\end{eqnarray}
If we had alternatively taken the mass of the new gauge boson to be small relative to the characteristic scale of momentum transfer, we would not be able to integrate out the mediator and a strict EFT power counting would not be appropriate \cite{Fitzpatrick:2012ix}. However, the scattering in a direct detection experiment would differ only by inverse powers of momentum transfer. The operators that introduce dipole interactions through a light mediator couple directly to the photon field strength $F_{\mu \nu}$, and these are described by
\begin{eqnarray} \label{eq:photon-DM-ops2}
\cO_{\rm \chi, MD}^{\mu \nu} & = & \frac{g^{\rm MD}}{\Lambda} \bar \chi i \sigma^{\mu \nu}  \chi ,\\
\label{eq:photon-DM-ops3}
\cO_{\rm \chi, ED}^{\mu \nu} & = & \frac{g^{\rm ED}}{\Lambda} \bar \chi i \sigma^{\mu \nu} \gamma_5 \chi.
\end{eqnarray}
As stated above, the interaction operator for $\chi$ is determined by the dynamics of the $\psi$ fermion(s). The anapole current in \Eq{eq:photon-DM-ops} will arise if charged $\psi^\pm$ states condense to form a neutral Majorana state $\chi$ \cite{Bagnasco:1993st}. The dipole currents form if an electromagnetically neutral $\psi^0$ couples to an electromagnetically charged pair of partner $\psi^\pm$ particles (of appropriate spin) \cite{Weiner:2012gm}. The scale at which the charged $\psi$ states are integrated out is $\Lambda$.

The model in \Eq{eq:UV-model} is quite simple, but the different dark matter interaction operators in Eqs.~(\ref{eq:photon-DM-ops})--(\ref{eq:photon-DM-ops3}) lead to a rich assortment of momentum and velocity dependences in the nuclear scattering cross section, as described in detail in \cite{Gresham:2014vja, Gluscevic:2015sqa}. The momentum and velocity dependence that appear in the differential cross section are characterized by one or more ``responses'' that contain information about the zero--momentum--transfer normalization, the overall momentum or velocity dependence, and a form factor that describes the shape of the spectrum at higher momentum transfer and relative velocity. We list these operators in \Tab{tab:operators}, highlighting the overall momentum and velocity dependence that multiply the dominant response or responses (this is an abbreviated version of the more exhaustive table that appeared in \cite{Gluscevic:2015sqa}, using results of \cite{Gresham:2014vja, Gluscevic:2015sqa}). Terms in the third column of \Tab{tab:operators} that multiply different form factors are separated with a comma; more specifically, using the definition
\begin{equation}\label{eq:specform}
\frac{d \sigma_T}{d \ER}(\ER,v) = \sum_{(n,n')} \sum_{X} f_X^T(\ER,v) \mathcal{R}_X^{T,(n,n')}(y) \, , 
\end{equation}
where the two summations are over nucleon permutations $(n,n') \in  \left[ (p,p),(n,n),(p,n),(n,p) \right]$ and target-dependent nuclear response functions $X \in \left[M,\Sigma',\Sigma'',\Phi'',\Delta,M\Phi'',\Delta\Sigma' \right]$ (as defined in \cite{Anand:2013yka}), the terms in the third column of Table~\ref{tab:operators} illustrated the EFT dependence of the various $f_X^T(\ER,v)$. In the above definition of \Eq{eq:specform}, the variable $y \equiv m_T \ER b^2/2$, where $b \equiv \sqrt{41.467/(45A^{-1/3} - 25A^{-2/3})}$ fm is the harmonic oscillator parameter for an atom with atomic number $A$. In this work, we will focus on differentiating interactions that have the same momentum scaling but different velocity dependence.

\begin{table}[tb]
\begin{centering}
\renewcommand{\arraystretch}{2.0}
\begin{tabular}{|c||c|c|c|} \hline
 Model name & {\rm Lagrangian} & \text{$\vec q$, $v$ Dependence} & \text{$\sigma_p$} 
\\ \hline \hline
 SI & $\frac{g}{M^2}\bar{\chi} \chi \bar{N} N$ & 1 & $\frac{\mu_p^2}{\pi}\left(\frac{f_p}{M^2} \right)^2$
\\ \hline 
 Anapole & $\frac g{M^2}\bar{\chi} \gamma^\mu \gamma_5 \chi \, \cJ_\mu $ & $v_\perp^2, \vec{q}^{\, 2}/m_N^2  $ & $\frac{\mu_p^2}{\pi}\left(\frac{e g^{\text Anapole}}{M^2} \right)^2$
\\ \hline
Magnetic Dipole (heavy) & $\frac g{ M^2\Lambda} \bar{\chi} \sigma^{\mu \nu} \chi  \, q_\nu \cJ_\mu $ & $\frac{\vec q^{\,4}}{\Lambda^4}+ \frac{\vec{q}^2 v_\perp^2 }{\Lambda^2},\vec q^{\,4}/\Lambda^4$ & $\frac{\mu_p^2}{\pi}\left(\frac{e g^{\text MD}}{M^2} \right)^2 \frac{\vec{q}_{\text ref}^2}{\Lambda^2}$
\\ \hline
Electric Dipole (heavy) &$ \frac g{M^2 \Lambda} \bar{\chi} \sigma^{\mu \nu} \gamma_5 \chi \, q_\nu \cJ_\mu $ & $\vec{q}^2 /\Lambda^2 $ & $\frac{\mu_p^2}{\pi}\left(\frac{e g^{\text ED}}{M^2} \right)^2 \frac{\vec{q}_{\text ref}^2}{\Lambda^2}$
\\ 
\hline 
Magnetic Dipole (light) & $\frac{g}{\Lambda} \bar{\chi} \sigma^{\mu \nu} \chi F_{\mu\nu} $ & $1+ \frac{v_\perp^2 m_N^2}{\vec{q}^2 }$, 1  & $\frac{\mu_p^2}{\pi}\left(\frac{e g^{\text MD}}{\Lambda |\vec{q}_{\text ref}|} \right)^2 $
\\ \hline
Electric Dipole (light) & $\frac{g}{\Lambda} \bar{\chi} \sigma^{\mu \nu} \gamma_5 \chi F_{\mu\nu} $ & $m_N^2/\vec{q}^2 $ & $\frac{\mu_p^2}{\pi}\left(\frac{e g^{\text ED}}{\Lambda |\vec{q}_{\text ref}|} \right)^2 $
\\ \hline \hline
\end{tabular}
\caption{Test interaction models considered in this work, listed by name, Lagrangian, and definition of $\sigma_p$ in the first, second, and third column respectively. In the third column we list their associated momentum and velocity dependences (adapted from \cite{Gluscevic:2015sqa}). The labels `light' and `heavy' in the dipole models denote the magnitude of the mediator mass relative to the characteristic momentum transfer. The nucleon electromagnetic current $\cJ_\mu$ is defined in \Eq{eq:current}; the transverse velocity $v_\perp$ and momentum transfer $\vec q$ are discussed in \Sec{subsec:momentum_velocity}; $f_p$ is the proton coupling (here we take $f_p=f_n$, where $f_n$ is the neutron coupling); $\mu_p$ is the dark matter-proton reduced mass; $q_{\rm ref}$ is a reference momentum characterizing the `turn-over' of the energy spectrum, taken here to be $100$ MeV; and $\Lambda$ is a heavy mass or compositeness scale appearing in the dipole models. Terms in the third column that induce different nuclear responses, and thus require different form factors, are separated by a comma (see \eg \cite{Anand:2013yka} for more details). }
\label{tab:operators} 
\end{centering}
\end{table}

\subsection{Simulations\label{sec:sims}}
\begin{table*}[tbp]
  \setlength{\extrarowheight}{3pt}
  \setlength{\tabcolsep}{10pt}
  \begin{center}
	\begin{tabular}{|c||m{2.3cm}|m{4.2cm}|m{3.2cm}|}  \hline
	Label & A (Z) & Energy window [keVnr] & Exposure [kg--yr] \\ \hline
	\hline
	Xe & 131 (54) & 5--40 & 2000 \\  \hline
	Ge & 73 (32) & 0.3--100 & 100  \\  \hline
	F &  19 (9) & 3--100 & 606 \\  \hline
	\hline
	Xe(x3) & 131 (54) & 5--40 & 6000 \\  \hline
	Xe(x10) & 131 (54) & 5--40 & 20 000 \\  \hline
	XeG3 & 131 (54) & 5--40 & 40 000 \\ \hline \hline
	\end{tabular}
  \end{center}
\caption{Mock experiments considered in this work. The efficiency and the fiducialization of the target mass are included in the exposure. The first group of experiments is chosen such to be representative of the reach of G2 experiments for xenon, germanium, and fluorine targets. The exposure for xenon and germanium is chosen to agree with the projected exclusion curves for LZ and SuperCDMS presented in Ref.~\cite{Cushman:2013zza}. The second group of experiments is used to quantitatively assess the impact of including the timing information as a function of the exposure (\ie the observed number of events).  }
\label{tab:experiments}
\end{table*}
\begin{table*}[t] 
\setlength{\extrarowheight}{3pt}
\setlength{\tabcolsep}{12pt}
\begin{center}
\begin{tabular}{|c||m{3cm}|m{3cm}|m{3cm}|}\hline
Interaction /target & Xe & Ge & F\\
\hline\hline 
%$m_\chi$ [GeV] & (20, 125, 500) & (20, 125, 500) & (20, 125, 500) \\
%\hline\hline 
SI& (106, 100, 99)& (10, 4, 4)& (5, 1, 2)\\ \hline
Anapole& (110, 99, 98)& (12, 5, 6)& (39, 3, 3)\\ \hline
Mag. dip. heavy& (111, 90, 89)& (4, 5, 5)& (5, 1, 1)\\ \hline
Mag. dip. light& (108, 103, 103)& (36, 15, 15)& (90, 16, 16)\\ \hline
Elec. dip. heavy& (108, 92, 89)& (4, 4, 4)& (1, 0, 0)\\ \hline
Elec. dip. light& (106, 103, 102)& (63, 15, 14)& (41, 12, 12)\\ \hline

\end{tabular}
\end{center}
\caption{Predicted number of events in G2 experiments for various interactions with xenon, germanium, and fluorine targets assuming a DM mass of ($20$ GeV, $125$ GeV, and $500$ GeV), for a cross section set to the current upper limits. Labels `light' and `heavy' denote the relative relation between the mediator mass and the characteristic scale of momentum transfer. }
\label{tab:pred_events}
\end{table*}

For our simulations, we consider the interactions discussed in the previous Section (summarized in Table~\ref{tab:operators}), for three benchmark DM particle masses: $20$ GeV, $125$ GeV, and $500$. Furthermore, we optimistically set the cross sections to be the value maximally allowed by LUX~\cite{Akerib:2016vxi} \footnote{We note that LUX currently produces the most constraining bound on the models and masses considered in this paper, although the constraint from PandaX--II is only marginally weaker~\cite{Tan:2016zwf}.}. Our baseline analysis focuses on G2 experiments employing xenon, germanium, and fluorine targets. Since fluorine experiments measure only the energy--integrated rate, we assume that fluorine has no energy resolution. For the rest of the experiments, we assume a perfect energy resolution, which should be a good approximation for our purposes \cite{Gluscevic:2015sqa}. The exposure and energy window of our mock experiments are summarized in Table~\ref{tab:experiments}. Throughout the analysis, we assume unit detection efficiency and zero backgrounds. In addition to the aforementioned, we also consider the potential reach of a Generation 3 (G3) xenon experiment, as well as various xenon experiments with exposures lying somewhere between G2 and G3 (the properties of which are summarized in Table~\ref{tab:experiments}).  We define G3 to be the experiments reaching the neutrino floor \cite{Billard:2013qya}. The predicted number of events for each interaction considered in these mock experiments are shown in Table~\ref{tab:pred_events}. 

Each simulated recoil data set is generated by randomly selecting from a Poisson distribution with a mean given by the predicted number of events; the predicted number of events is calculated using \Eq{eq:totevents}, using the astrophysical parameters listed in \Sec{sec:rate} and incorporating the effect of gravitational focusing by the Sun following the procedure of Ref.~\cite{Lee:2013wza}. The recoil energy and time of each event is then obtained by applying a rejection sampling algorithm to the two--dimensional differential scattering rate. This procedure is repeated for $\cO(50)$ simulations in order to assess the variability of results arising from Poisson noise. 

\subsection{Analysis method}\label{sec:stats}

We analyze each simulated data set using Bayesian inference framework. In this framework, the probability that the data $\vec{X}$ assigns to a given model $\cM_j$ is given by
\begin{equation}\label{eq:probs}
P(\cM_j) = \frac{\cE_j(\vec{X}|\cM_j)}{\sum_i \cE_i(\vec{X}|\cM_i)} \, ,
\end{equation}
where the sum is performed over all competing hypotheses, and $\cE(\vec{X}|\cM)$ is the evidence of model $\cM$, defined as
\begin{equation}\label{eq:evidence}
\cE(\vec{X}|\cM) = \int d\Theta \, \cL(\vec{X}|\Theta,\cM) \, p(\Theta,\cM) \, ,
\end{equation}
and is intuitively understood to be the factor required to normalize the posterior probability distribution $\cP$,
\begin{equation}\label{eq:posterior}
\cP(\Theta | \vec{X}, \cM) = \frac{\cL(\vec{X}|\Theta,\cM)\, p(\Theta,\cM)}{\cE(\vec{X}|\cM)} \, . 
\end{equation}
Here, $\cL(\vec{X}|\Theta,\cM)$ is the likelihood, \ie the probability of obtaining the data, given a particular model $\cM$ and parameters $\Theta$ (for the purpose of this analysis $\Theta = \cbL m_\chi, \sigma_p \cbR$), and $p(\Theta, \cM)$ is the prior. In order to remain as agnostic as possible, we take wide priors in both $m_\chi$ and $\sigma_p$\footnote{Log priors are taken for both $m_\chi$ and $\sigma_p$, spanning $1--3000$ GeV in mass and $7$ orders of magnitude in cross section.}. We use an unbinned likelihood function of the form
\begin{equation}\label{eq:likelihood}
\cL(\vec{X}|\Theta,\cM) = \frac{\left<N \right>^N}{N!} \, e^{-\left<N \right>} \, \prod_{x_i \in \vec{X}}\, \frac{1}{\left<N \right>} \, \frac{dR}{d\ER dt} \bigg|_{\ER,t \, = \, x_i} \, ,
\end{equation}
where $\left<N \right>$ is the predicted number of events, $N$ is the number of observed events, and the product runs over all observed events, with a recorded energy and time label $x_i \equiv \cbL E_{R,i} \, , \, t_i \cbR$. When time (or $\ER$) information is neglected, the differential rate is taken to be averaged over that variable. 

For each of our simulated recoil data sets, we use a nested sampling algorithm implemented in \texttt{MultiNest} software package \cite{pymultinest,Feroz:2008xx,Feroz:2007kg,Feroz:2013hea} to reconstruct the posterior\footnote{\texttt{MultiNest} runs are performed with 2000 live points, an evidence tolerance of 0.1, and a sampling efficiency of 0.3.}. Once we compute the evidence of all competing models, we estimate the probability of successfully identifying the true model using \Eq{eq:probs}. This procedure is then repeated for $\cO(50)$ simulations to assess the probability of successful model identification (the variability arises from Poisson fluctuations). A model is correctly identified if the probability determined using \Eq{eq:probs} is large. For the purpose of this paper, we define identify a ``successful'' model selection with an outcome: $P \geq 90\%$. The primary quantity of interest for future direct detection experiments is then the fraction of simulations which lead to a successful model identification, which we refer to as the ``success rate'' in the following Section.    

To evaluate the success rate, we apply kernel density estimation (KDE) with a Gaussian kernel to an ensemble of posterior probabilities a simulated data set assigns to the true underlying model. In the following Section, we will show the KDE distribution for each experimental combination (derived from both with-- and without--time analyses), and determine the success rate by integrating the distributions above the 90\% threshold.

\section{Results}\label{sec:results}
\begin{figure}
\centering
\includegraphics[width=0.4\textwidth, trim=.6cm 0.0cm .6cm 0.0cm,clip=true]{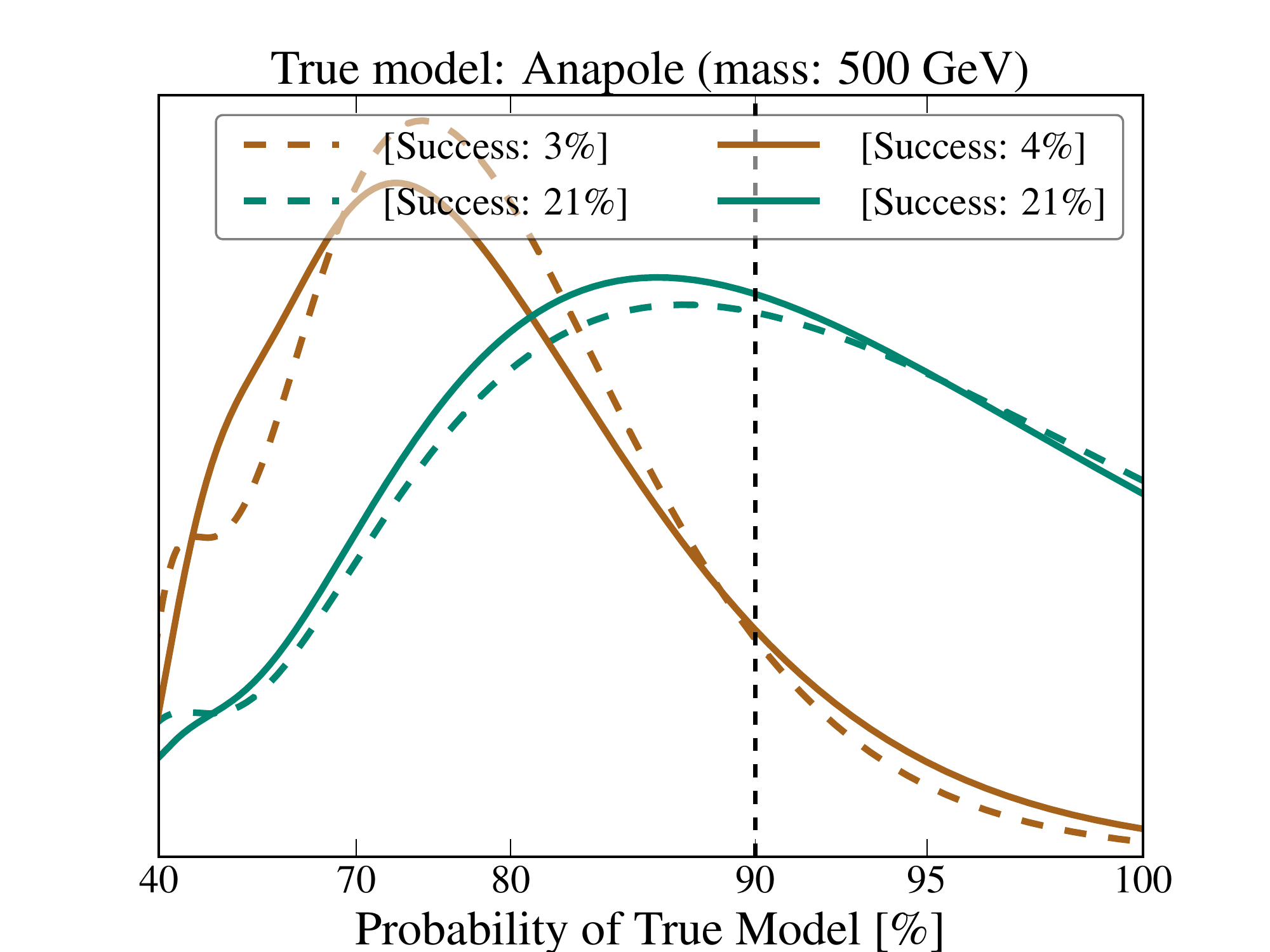}
\includegraphics[width=0.4\textwidth, trim=.6cm 0.0cm .6cm 0.0cm,clip=true]{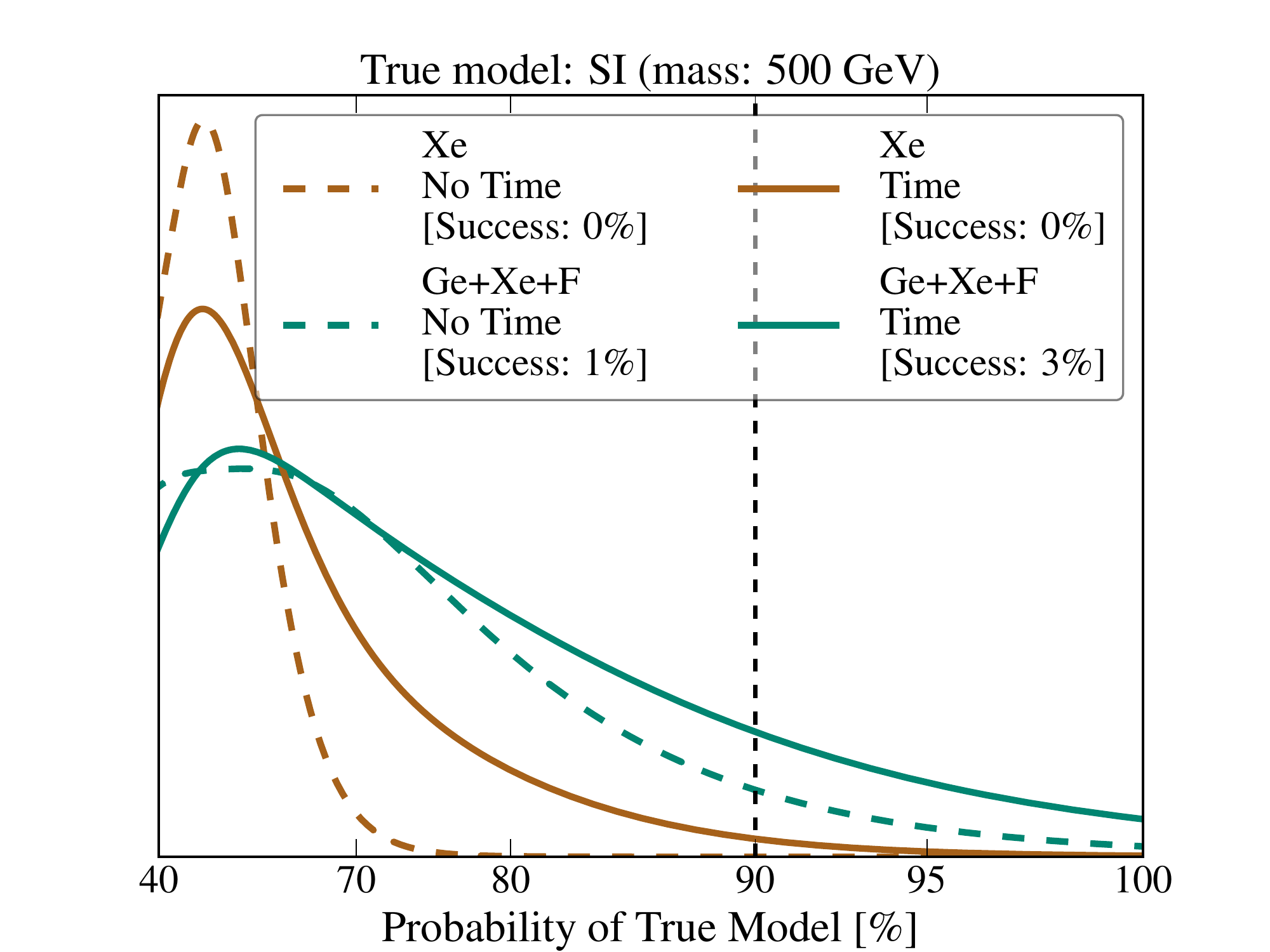}

\includegraphics[width=0.4\textwidth, trim=.6cm 0.0cm .6cm 0.0cm,clip=true]{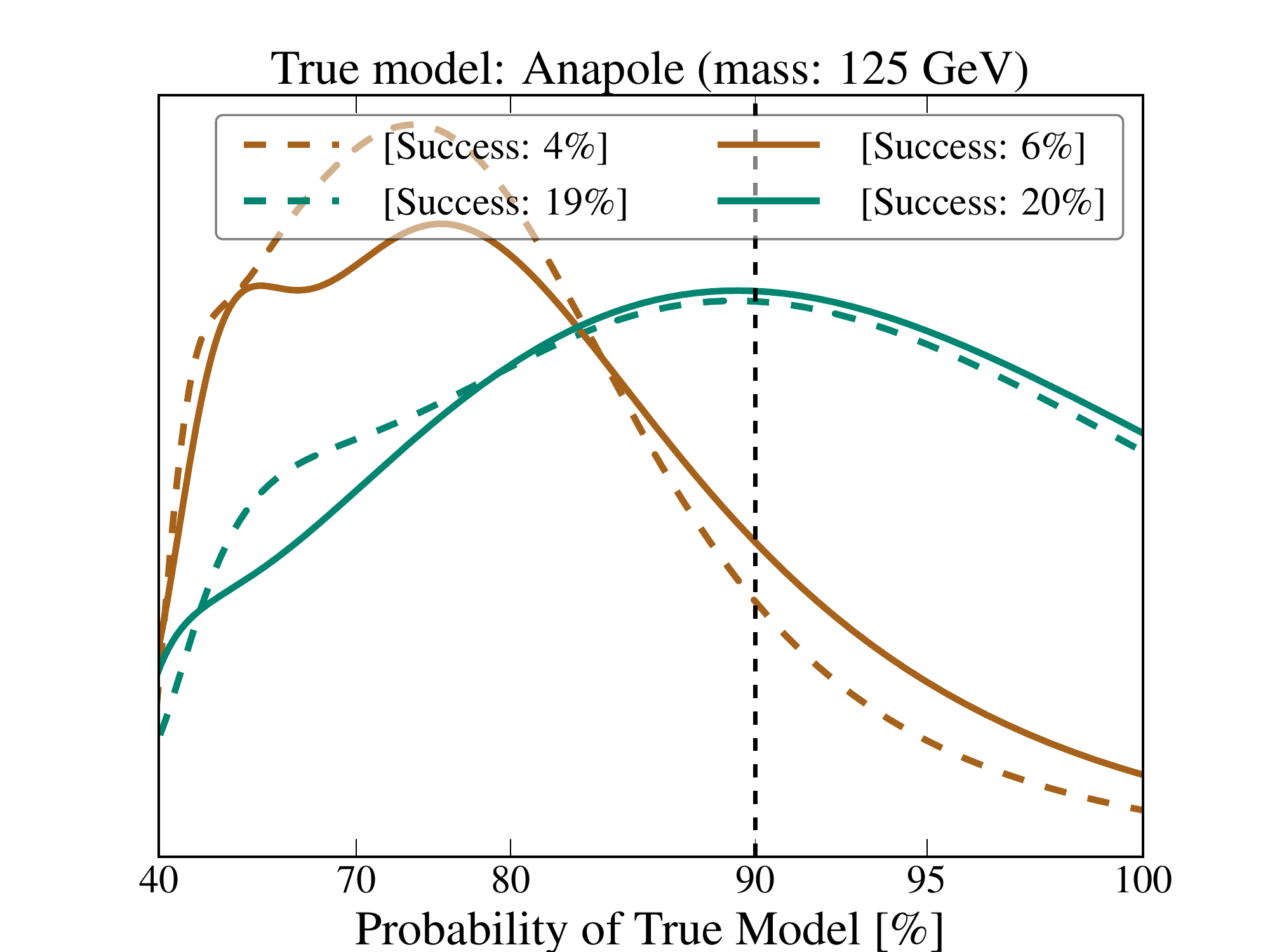}
\includegraphics[width=0.4\textwidth, trim=.6cm 0.0cm .6cm 0.0cm,clip=true]{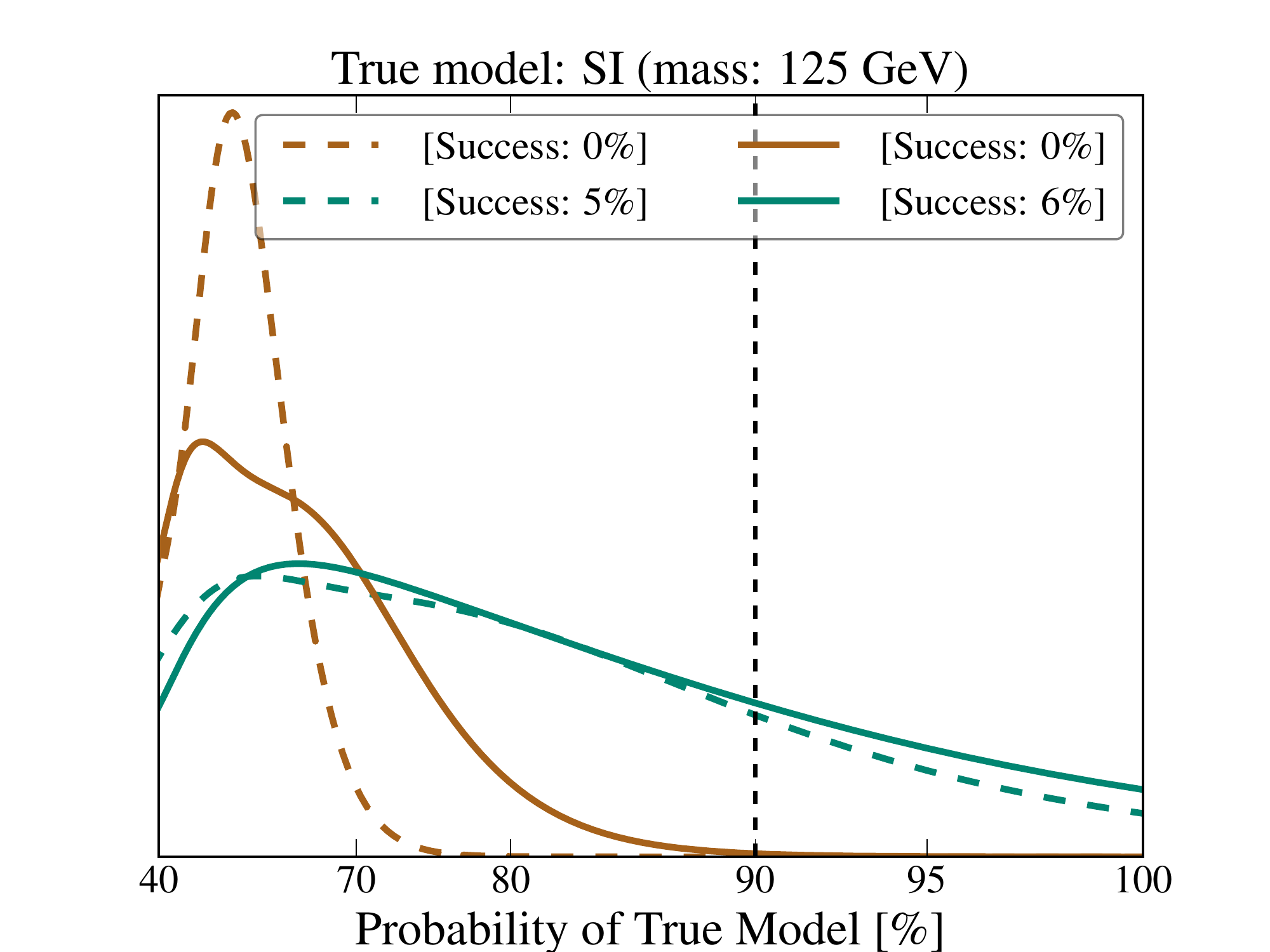}

\includegraphics[width=0.4\textwidth, trim=.6cm 0.0cm .6cm 0.0cm,clip=true]{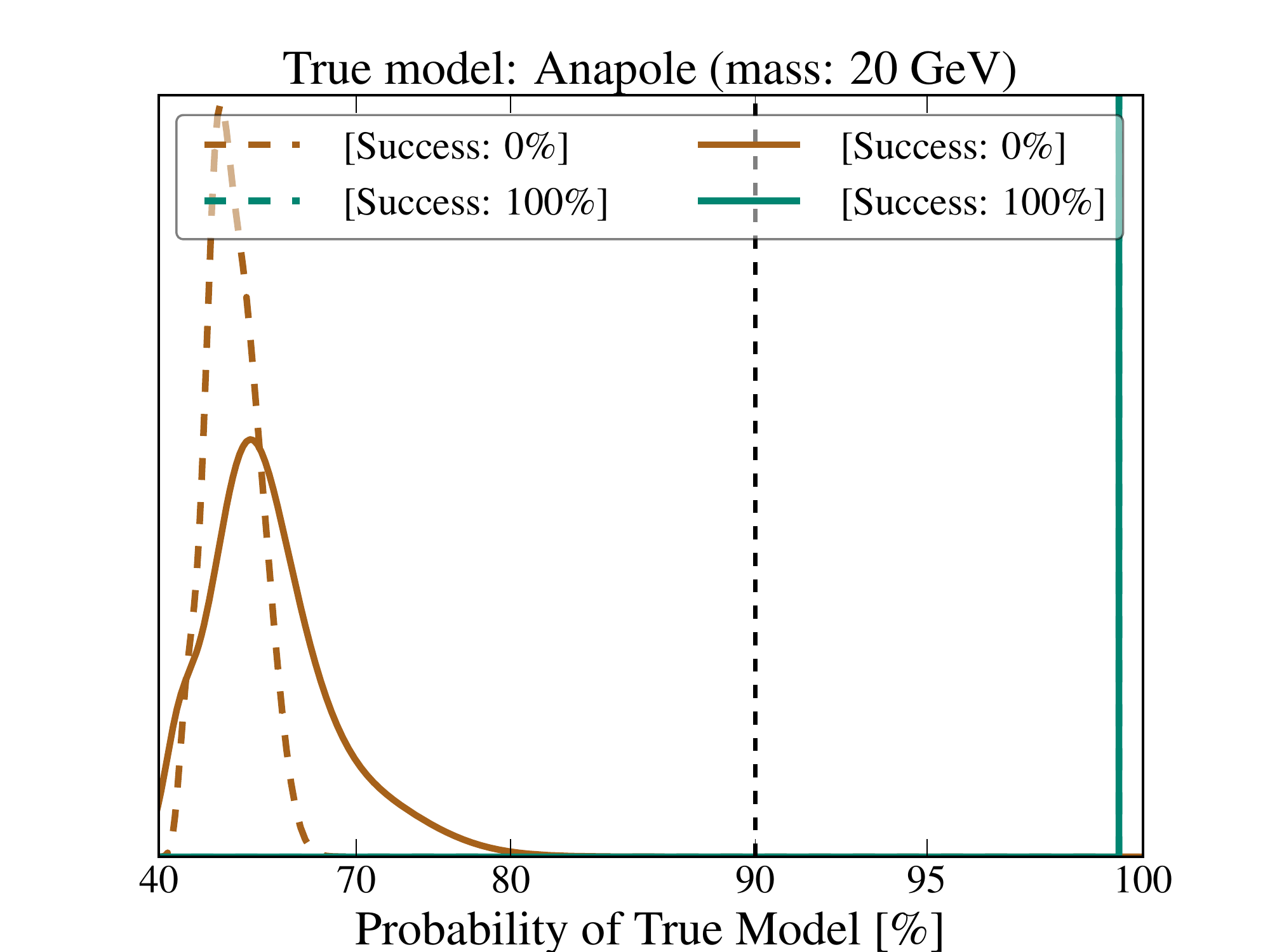}
\includegraphics[width=0.4\textwidth, trim=.6cm 0.0cm .6cm 0.0cm,clip=true]{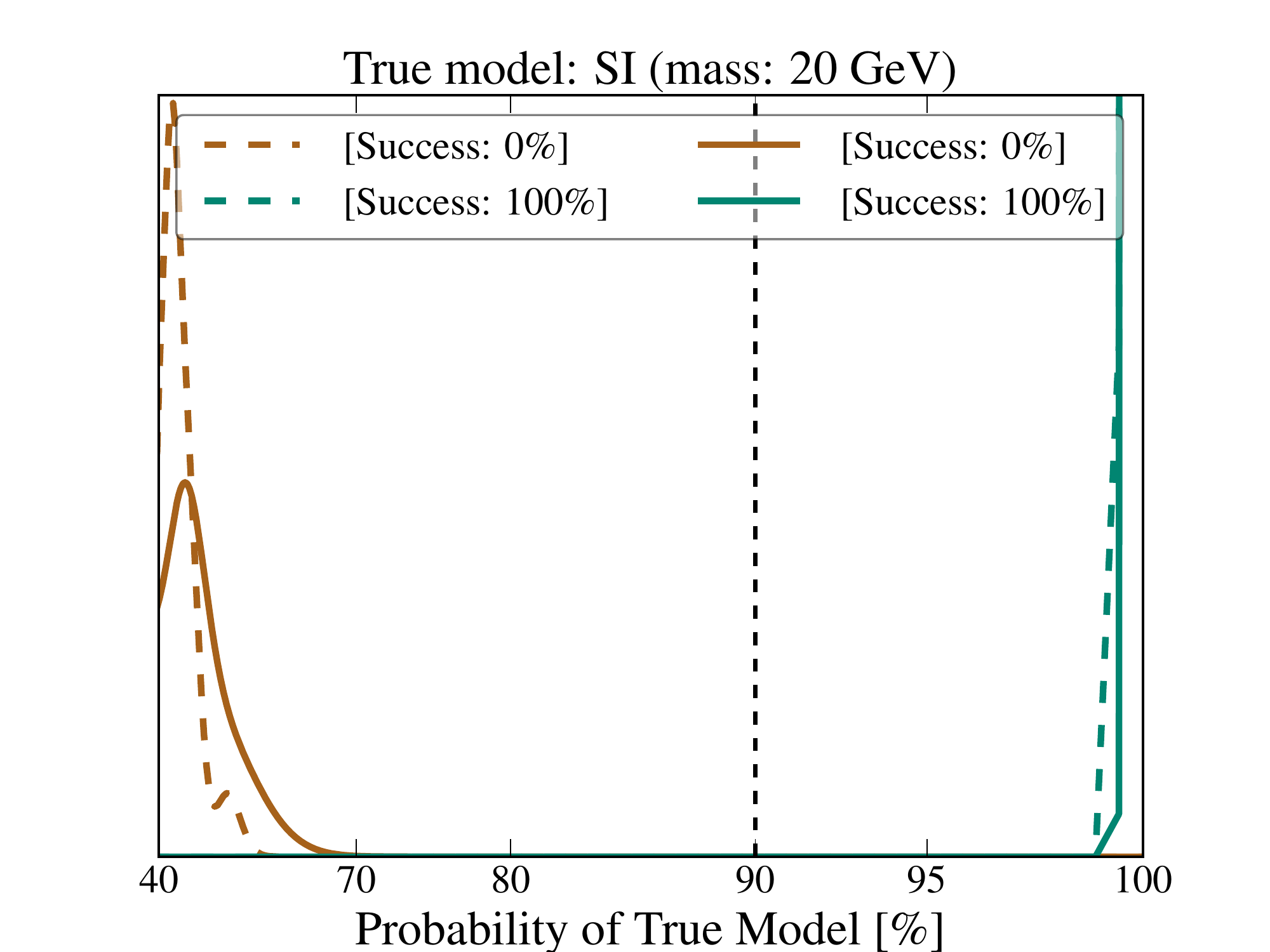}
\caption{\label{fig:gen2}
Model selection prospects with complimentary G2 targets. The reconstructed PDFs for the posterior probability assigned to the true underlying interaction model are shown for the anapole (left column) and SI (right column) interactions, for a $500$ GeV (top row), $125$ GeV (middle row), and $20$ GeV (bottom row) DM particle mass. Cross sections are set to current upper limits, and the PDFs are all normalized to unity between 0 and 100$\%$. Results are shown for the analyses of simulated data from a xenon experiment alone, and for a combined likelihood analyses of xenon, germanium, and fluorine experiments. Solid lines are obtained from analyses that include time information of the recoil events, while dashed lines are from those that do not. Success rate displayed in the legend represents the fraction of simulations for which the correct model was assigned $ \geq 90\%$ posterior probability (denoted by the vertical dashed line).}
\end{figure}

We first examine the extent to which including time information in the analysis of G2 experiments can help break degeneracy between models with the same momentum dependence, using as a case study the SI and anapole interactions. For this purpose, we simulate future G2 data for the SI and anapole interactions, and fit each simulation with these two models. We then compare the Bayesian evidences for the two models to evaluate the probability of the true underlying model (used to create a given simulation ensemble), as defined in \Eq{eq:probs}. We then derive the probability distribution function (PDF) of all possible outcomes (\ie a PDF of probabilities for identifying the true underlying model) from an ensemble of 50 simulations that had the same input model and parameters. Since we only consider two competing models, a PDF peaked around $50\%$ means that the data is most likely to be agnostic between the two models, \ie both models fit the data equally well\footnote{For this reason we never plot the low probability region of the PDFs}; this is the most pessimistic outcome possible. Conversely, a tail of the distribution at high probabilities, or a PDF shift in that direction, signifies improved model identification. \Fig{fig:gen2} shows the results of this exercise for likelihood analysis only for xenon simulations, and for a joint likelihood analysis performed on a combination of data obtained on xenon, germanium, and fluorine experiments.  DM particle masses used in the simulations are: $500$ GeV, $125$ GeV, and $20$ GeV, with cross sections set to their respective current upper limits. We show the results obtained both without taking into account time dependence of the signals, and including the modulation analysis.

Consistent with the results of~\cite{Gluscevic:2015sqa}, we find that the two models can be confidently distinguished for a signal close to the current detection threshold, provided G2--level exposure on xenon and a detection with a fluorine experiment, but only if data from these experiments are jointly analyzed (xenon and germanium experiments are not complementary in the sense that a joint analysis does not significantly improve prospects for model selection, and thus we do not display results for this case). For a low--mass DM particle (20 GeV), the improvement upon combining these two types of experiments is drastic: the PDF of possible model--selection outcomes entirely shifts to a delta--function at 100$\%$ probability in favor of the correct model. For intermediate and high masses, the prospects are still visibly improved, but not very optimistic (at best on the level of $\sim$20$\%$ success rate), mostly due to the reduced scattering rate on fluorine.  

\begin{figure*}
\centering
\includegraphics[width=0.7\textwidth]{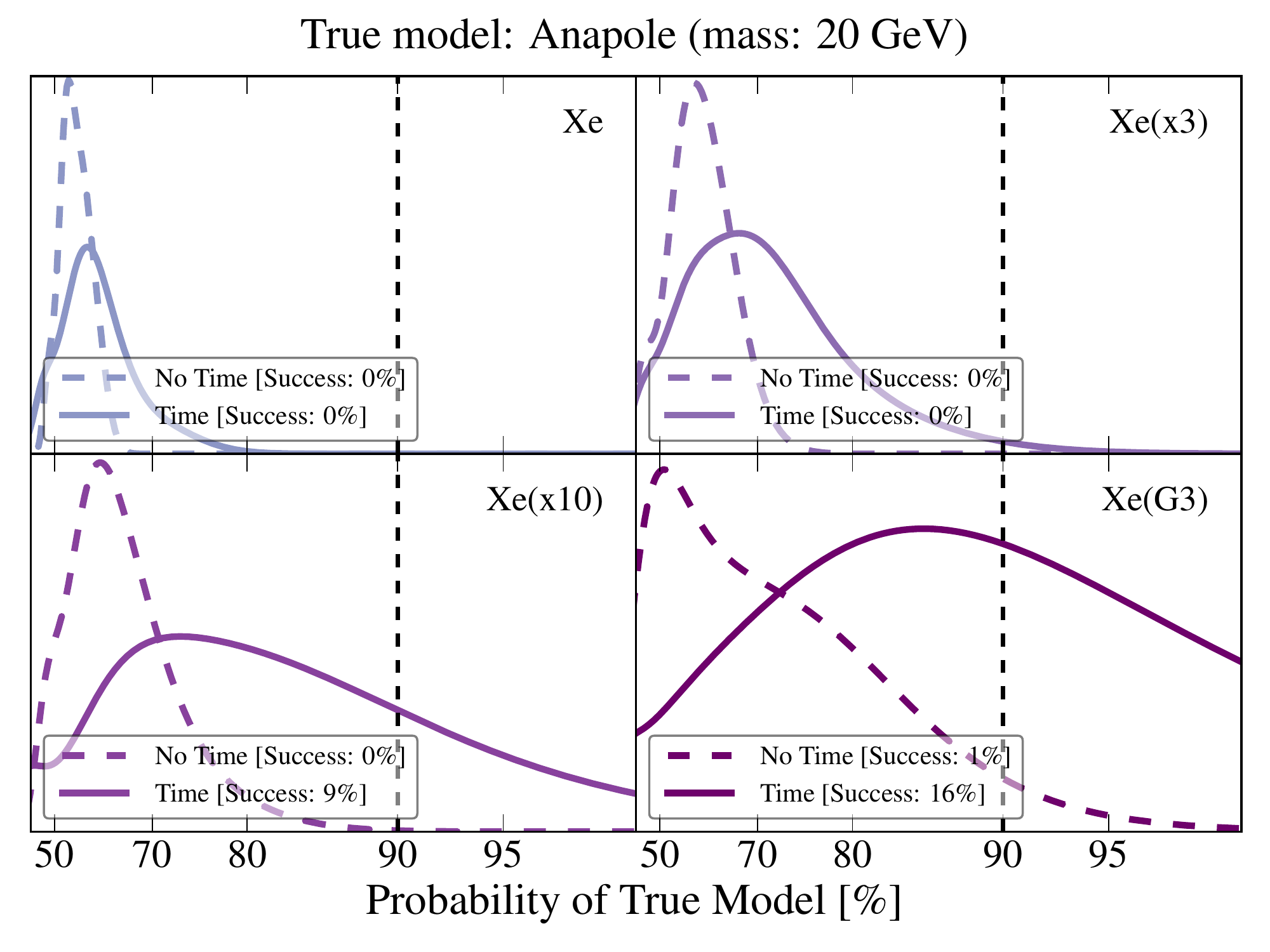}
\caption{\label{fig:20gev_anapole_XeFull_TNT_GF}
Model selection prospects for a single target (xenon), including (solid lines) and neglecting (dashed lines) time information in the likelihood analysis. The normalized PDFs are plotted for the probability of identifying the underlying model, here taken to be 20 GeV anapole DM, with a cross section that saturates the current upper limits. Panels from left to right, top to bottom, correspond to experimental exposures of 2, 6, 20, and 40 ton--years, respectively. Success rate displayed in the legend represents the fraction of simulations for which the correct model was assigned $ \geq 90\%$ posterior probability (denoted by the vertical dashed line).}

\end{figure*}
\begin{figure*}
\centering
\includegraphics[width=0.7\textwidth]{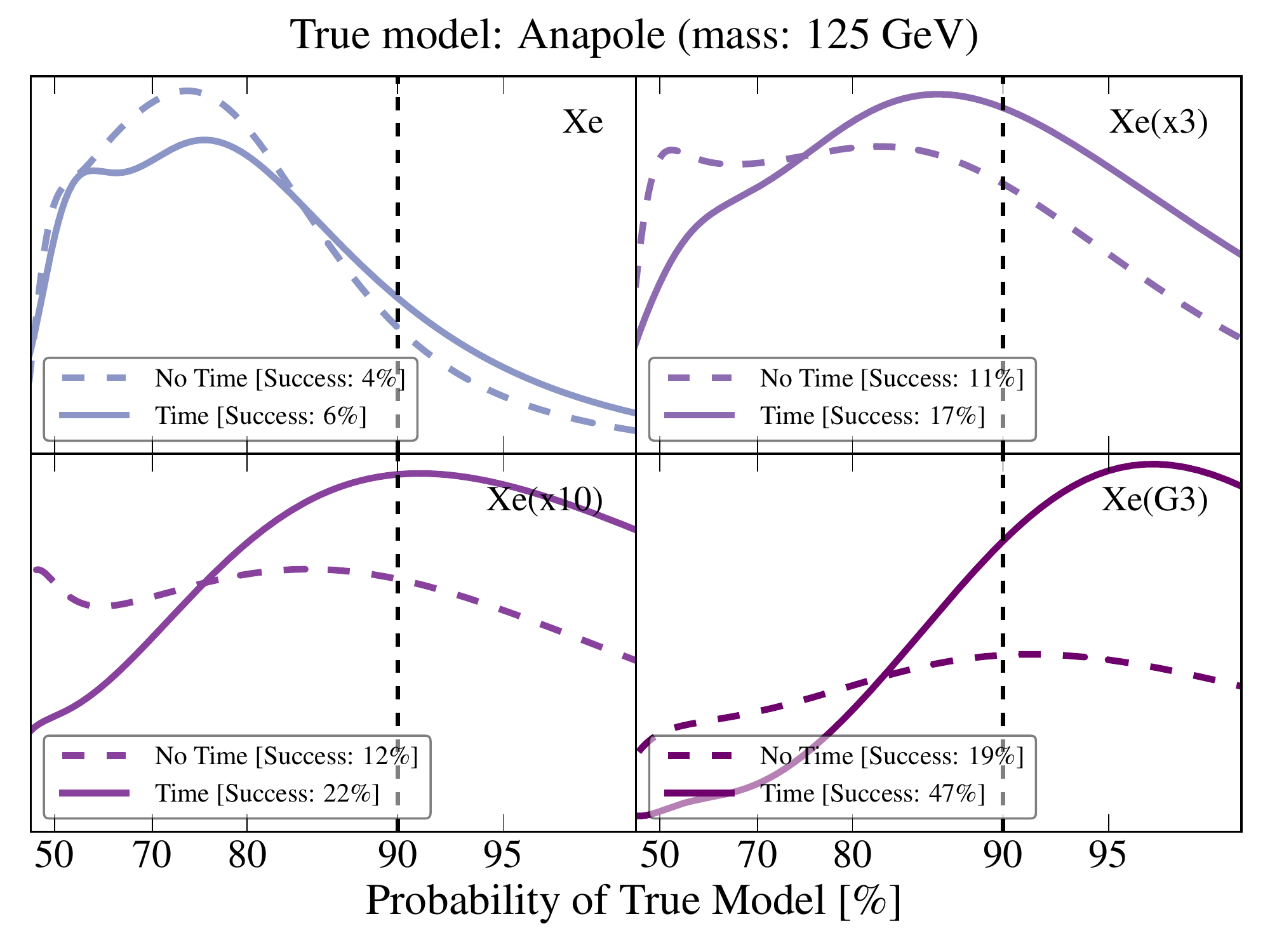}
\caption{\label{fig:125gev_anapole_XeFull_TNT_GF}
Same as Fig.~\ref{fig:20gev_anapole_XeFull_TNT_GF} but for $125$ GeV DM.}
\end{figure*}
\begin{figure*}
\centering
\includegraphics[width=0.7\textwidth]{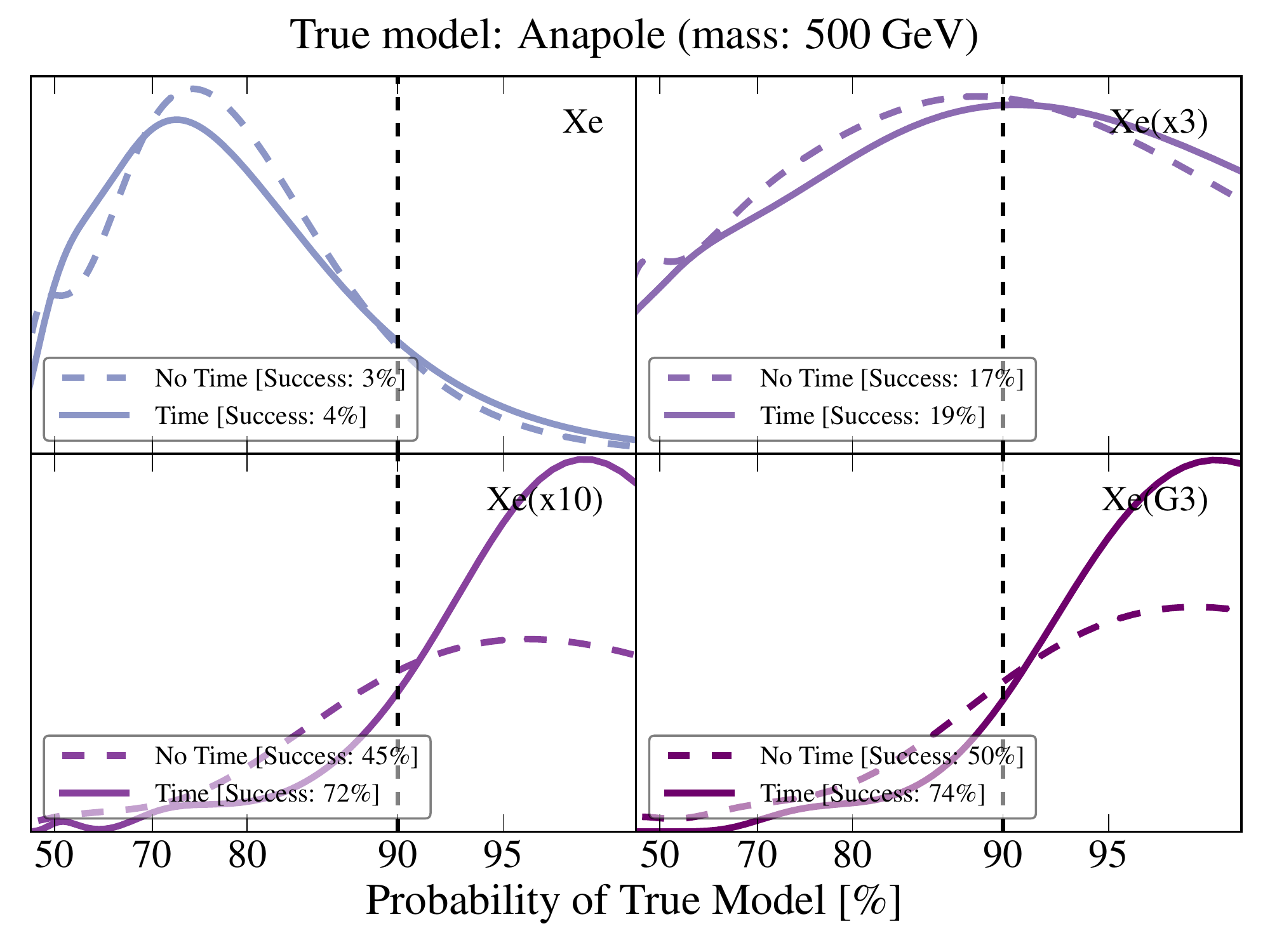}
\caption{\label{fig:500gev_anapole_XeFull_TNT_GF}
Same as Fig.~\ref{fig:20gev_anapole_XeFull_TNT_GF} but for $500$ GeV DM.}
\end{figure*}
\begin{figure*}
\centering
\includegraphics[width=0.7\textwidth]{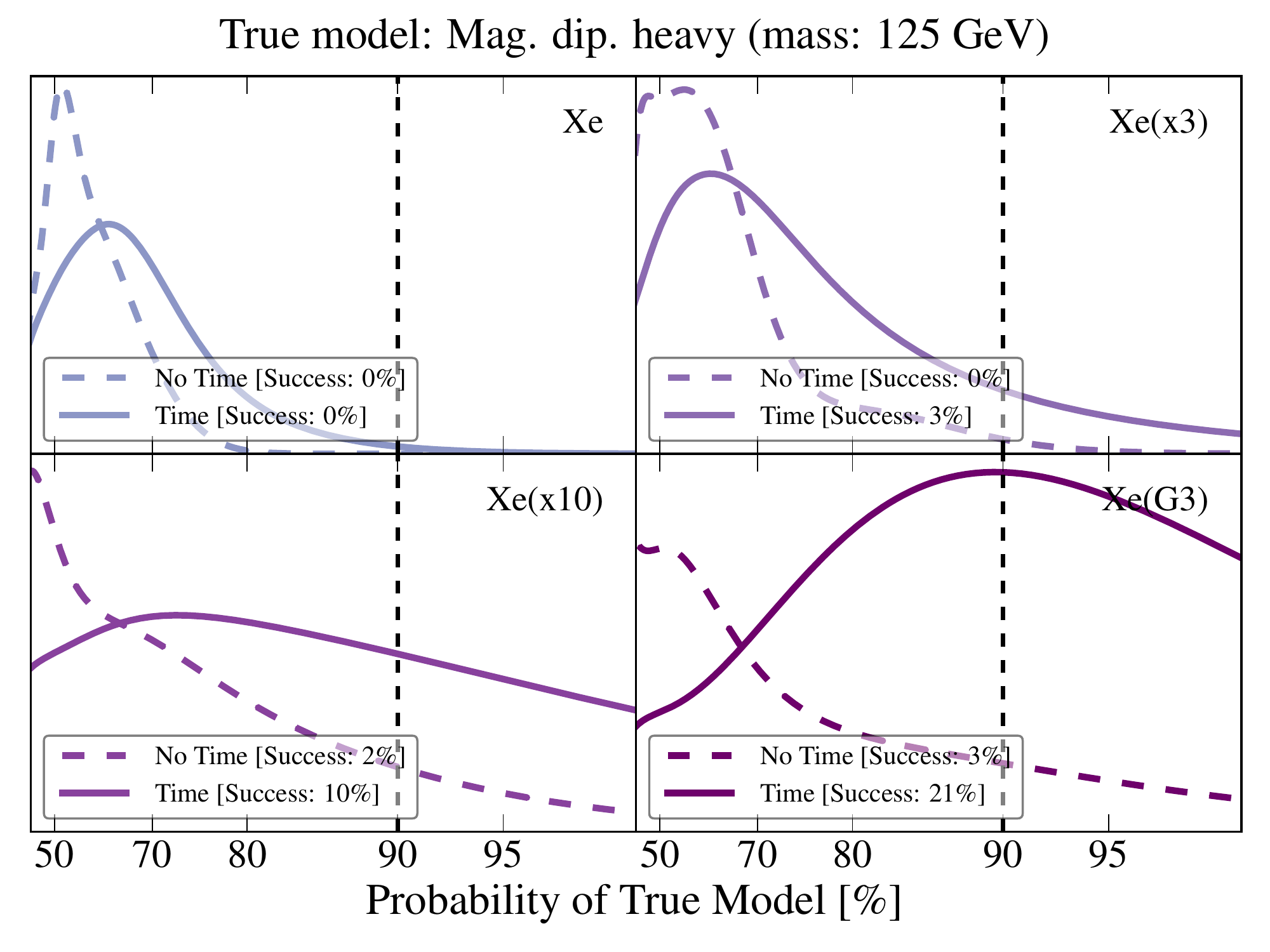}
\caption{\label{fig:125gev_Mag.dip.heavy_XeFull_TNT_GF}
Same as Fig.~\ref{fig:20gev_anapole_XeFull_TNT_GF}, but now assessing the ability of xenon experiments to break the degeneracy of the magnetic dipole (heavy mediator) and electric dipole (heavy mediator) interactions, instead of SI and anapole interactions. Simulations assume a $125$ GeV DM particle and a magnetic--dipole interaction.}
\end{figure*}
\begin{figure*}
\centering
\includegraphics[width=0.7\textwidth]{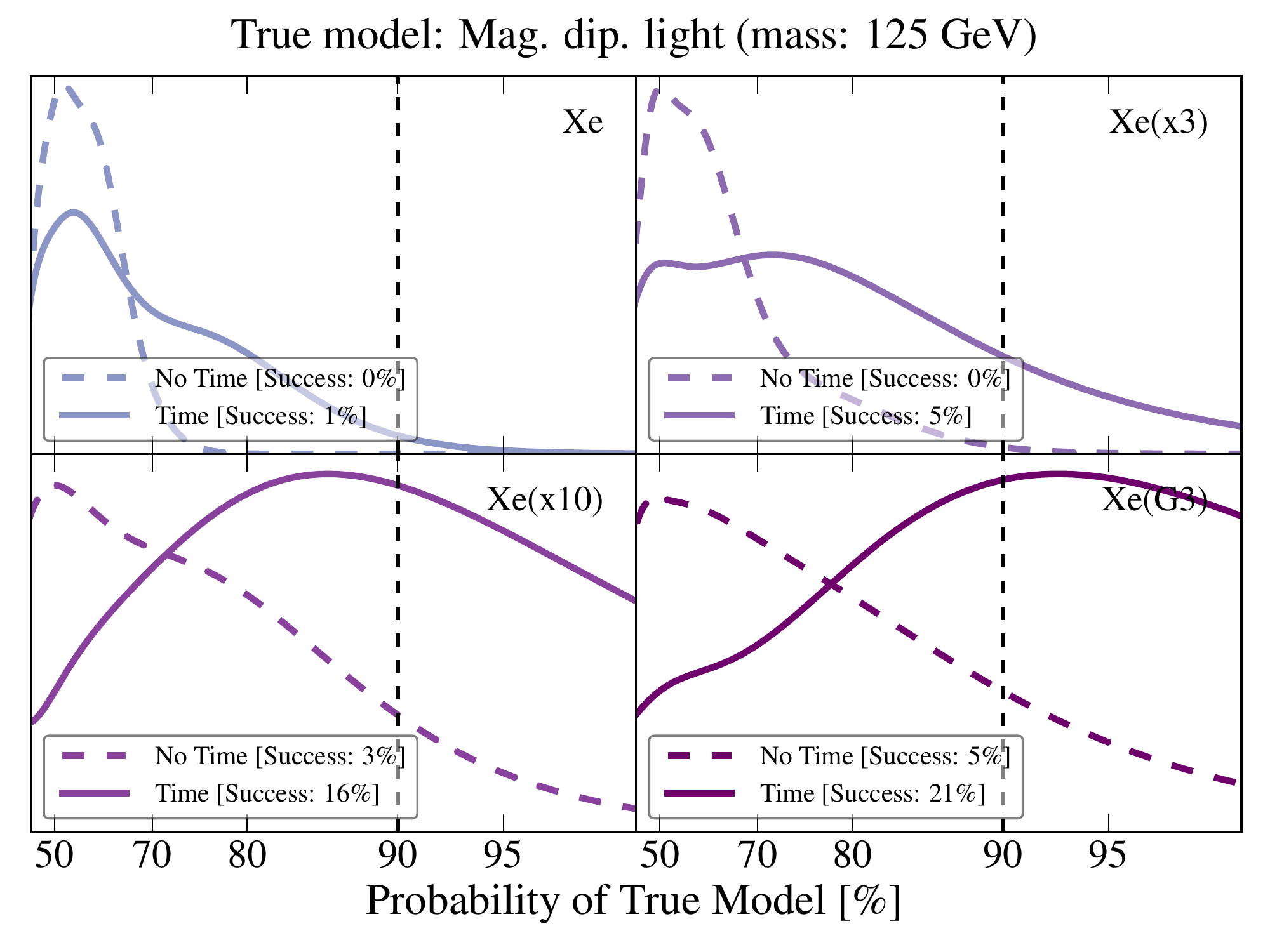}
\caption{\label{fig:125gev_Mag.dip.light_XeFull_TNT_GF}
Same as \Fig{fig:125gev_Mag.dip.heavy_XeFull_TNT_GF} but for a light--mediator case. }
\end{figure*}
Comparison of no--time and with--time analyses (displayed in dashed and solid lines, respectively) demonstrates that inclusion of time information only negligibly changes model--selection prospects for G2 level of exposures. Given that G2 experiments will optimistically detect on the order of $\simeq 100$ events, the statistical sample will be insufficient to clearly detect differences in the modulation signal that would otherwise aid differentiation between the two interactions.\footnote{See Sec.~4 of \cite{DelNobile:2015nua} for an estimation of the number of events needed for phase measurement. }

The question then arises of how many events are needed before the inclusion of time information can significantly help model selection prospects. We address this question in Figs.~\ref{fig:20gev_anapole_XeFull_TNT_GF}--\ref{fig:500gev_anapole_XeFull_TNT_GF} for a xenon experiment\footnote{We choose to illustrate this point on a xenon target, as xenon experiments are projected to observe far more events than their germanium or fluorine counterparts.}, showing the prospects (PDF of possible analysis outcomes) of successful model selection given a range of exposures: a 2, 6, 20, and 40 ton--year. Simulations used for these Figures all assume the anapole interaction and a DM mass of 20, 125, and 500 GeV, respectively. Results are shown for analyses that neglect (dashed) and include (solid) the time dependence of the rate. Simulations using the SI interaction are qualitatively similar and we defer these results to Appendix A.

From these Figures, we can infer that the addition of time information drastically improves prospects for successful model selection in the case of light DM particle (the solid--line PDFs are substantially shifted to the right in \Fig{fig:20gev_anapole_XeFull_TNT_GF}). In spite of this, even a G3 xenon experiment has only a $\simeq 16\%$ chance of disentangling these two interactions with high confidence. \Fig{fig:diff_rate_comp} gives an intuition for interpreting this outcome: both the recoil spectra and the phase of the modulation of the SI and anapole interactions (measured on a xenon experiment) are more degenerate for light than for heavy DM particles; thus, even when including time information in the analysis, a single--target experiment still must observe a large number of events in order to successfully distinguish between these models.

For larger DM masses, Figs.~\ref{fig:125gev_anapole_XeFull_TNT_GF} and \ref{fig:500gev_anapole_XeFull_TNT_GF} show better model--discrimination prospects at a fixed exposure, particularly when time is included in the analysis. From these Figures, we see that including time in the analysis can improve model selection in G3 experiments by as much as $\simeq 30\%$ for heavy DM, where the phases of the modulation signal for these two interactions may be misaligned by as much as $\simeq 5$ months (see again \Fig{fig:diff_rate_comp}). Finally, it is important to keep in mind that all the results displayed here assume the most optimistic number of observed recoil events. Thus, despite the improvement in model selection obtained with the inclusion of time information, it is likely that model identification will still be challenging using a single target, even with G3 experiments; most likely, the experiments will need to exploit both target complimentarily and the annual modulation in order to fully break the degeneracy between these types of models and ensure the highest chance of correctly identifying the interaction at hand.

The goal of this work was primarily a quantitative assessment of whether time information can be exploited in future direct detection analyses to break degeneracies in the recoil spectra of different interaction models --- and SI and anapole interactions provided a particularly illuminating case study for this purpose. However, the main conclusions presented in this Section hold for other sets of interactions as well, and we now briefly illustrate this point. In \Figs{fig:125gev_Mag.dip.heavy_XeFull_TNT_GF}{fig:125gev_Mag.dip.light_XeFull_TNT_GF} we consider a comparison of the magnetic dipole and electric dipole interactions for a $125$ GeV DM particle, assuming a heavy and light mediator, respectively. As before, we consider putative detections in future xenon experiments with exposures varying from 2 to 40 ton--years. The results are rather similar to the SI and anapole comparison in that G3 experiments can expect a $\simeq 20\%$ improvement in model selection when time is included in the analysis, but again necessitate target complementarity to have a high chance of confidently differentiating these interactions.

%%%%%%%%%%%%%%%%%%
\section{Summary and Discussion}\label{sec:conclusion}
We have considered here the potential impact of using time information in the likelihood analysis of data from future direct detection experiments in order to break degeneracies between the recoil energy spectra of different dark matter--baryon interactions. Specifically, we performed a statistical assessment of the prospects for successful Bayesian model selection between different interactions, using an ensemble of simulations that account for the impact of Poisson fluctuations. As a case study, we compared the standard spin--independent interaction and anapole dark matter, but also demonstrated that the main findings hold for other degenerate interaction models as well. We explored three different dark matter masses, and focused specifically on the most optimistic case where the cross sections for all interactions saturate the current upper limits. 

We found that even under the most optimistic of circumstances, including time information in the analysis of Generation 2 direct detection experiments does not significantly improve prospects for distinguishing between models with degenerate recoil spectra. Rather, correct model identification in Generation 2 experiments will almost certainly require measurements and combined analyses on multiple target elements. We found that for the inclusion of time information to significantly increase chances for successful model selection (by $\cO(10)\%$), for observations in an experiment employing a single target element, $\cO(1000)$ and $\cO(500)$ events must be observed for a 20 GeV and a 500 GeV DM particle respectively. These numbers are consistent with the `back--of--the--envelope' calculations performed in \cite{DelNobile:2015nua}. Furthermore, even if time information is exploited in Generation 3 xenon experiments, target complementarity must also be exploited to unequivocally differentiate between interaction models. We emphasize again that this finding holds even for the maximally optimistic scenario in which the interaction cross sections are as large as possible; it will be even more relevant for the case where the signals do not saturate current upper bounds.

In the event of a putative signal, direct detection experiments will be charged with the difficult task of illuminating the high energy behavior of dark matter solely from the observed low--energy recoils. This is a particularly daunting task in light of the fact that many feasible models produce nearly degenerate recoil spectra. Exploiting all of the information available, including the time dependence of nuclear recoil events explored in this work, will be necessary to make definitive statements regarding the true particle nature of dark matter. 
\bigskip

\textbf{Acknowledgments.} SW is supported under the University Research Association (URA) Visiting Scholars Award Program, and by a UCLA Dissertation Year Fellowship. VG gratefully acknowledges the support of the Schmidt Fellowship at the Institute for Advanced Study. SDM is supported by NSF PHY1316617. %Fermilab is operated by Fermi Research Alliance, LLC, under Contract No. DE-AC02-07CH11359 with the US Department of Energy. 

\appendix

\section{Model Selection Prospects in Xenon (SI Interaction)}
\begin{figure*}
\centering
\includegraphics[width=0.7\textwidth]{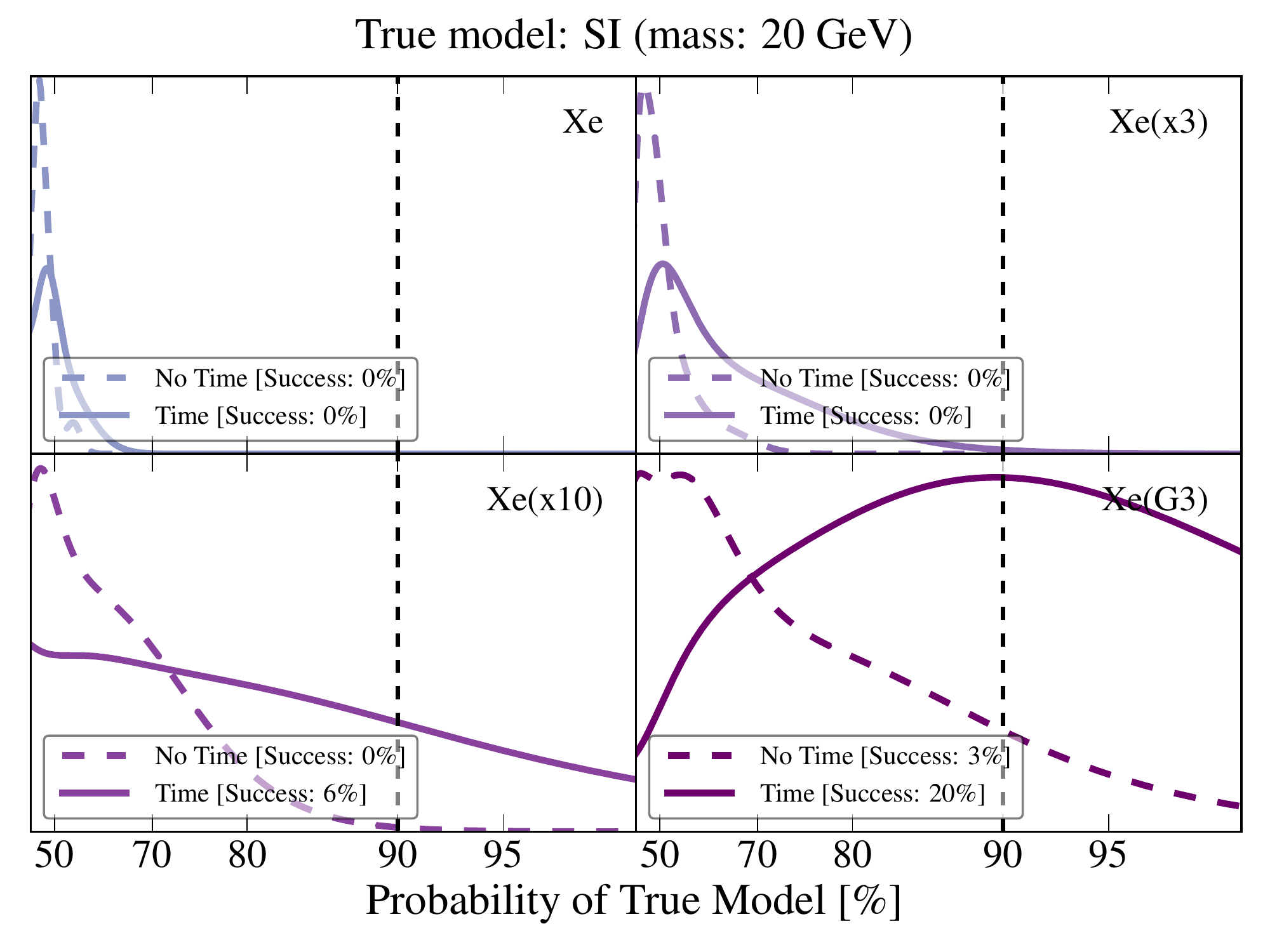}
\caption{\label{fig:20gev_SI_Higgs_XeFull_TNT_GF}
Same as Fig.~\ref{fig:20gev_anapole_XeFull_TNT_GF} but for the SI interaction as the true model.}
\end{figure*}

\begin{figure*}
\centering
\includegraphics[width=0.7\textwidth]{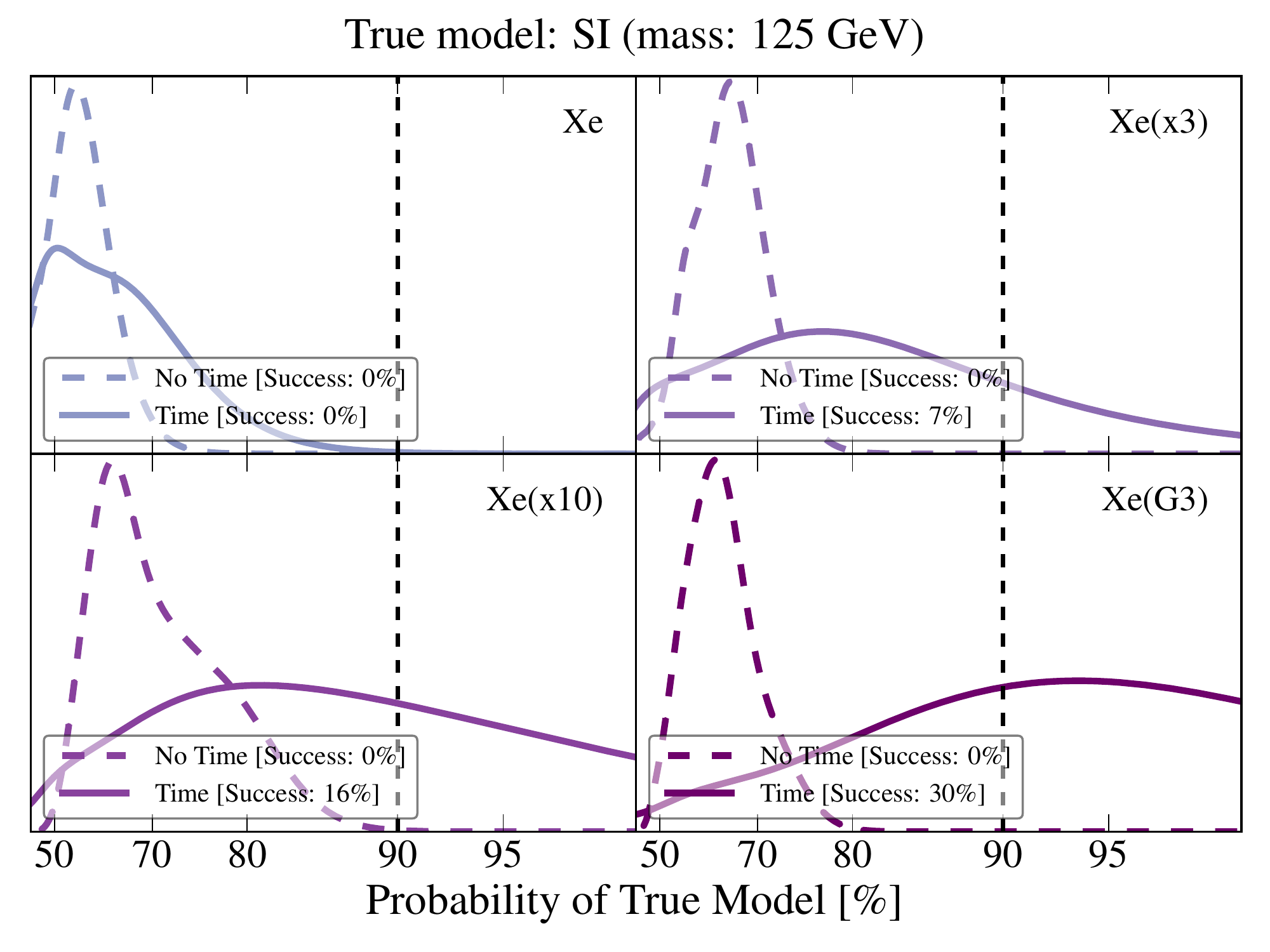}
\caption{\label{fig:125gev_SI_Higgs_XeFull_TNT_GF}
Same as Fig.~\ref{fig:20gev_anapole_XeFull_TNT_GF} but for a $125$ GeV DM and the SI interaction as the true model.}
\end{figure*}

\begin{figure*}
\centering
\includegraphics[width=0.7\textwidth]{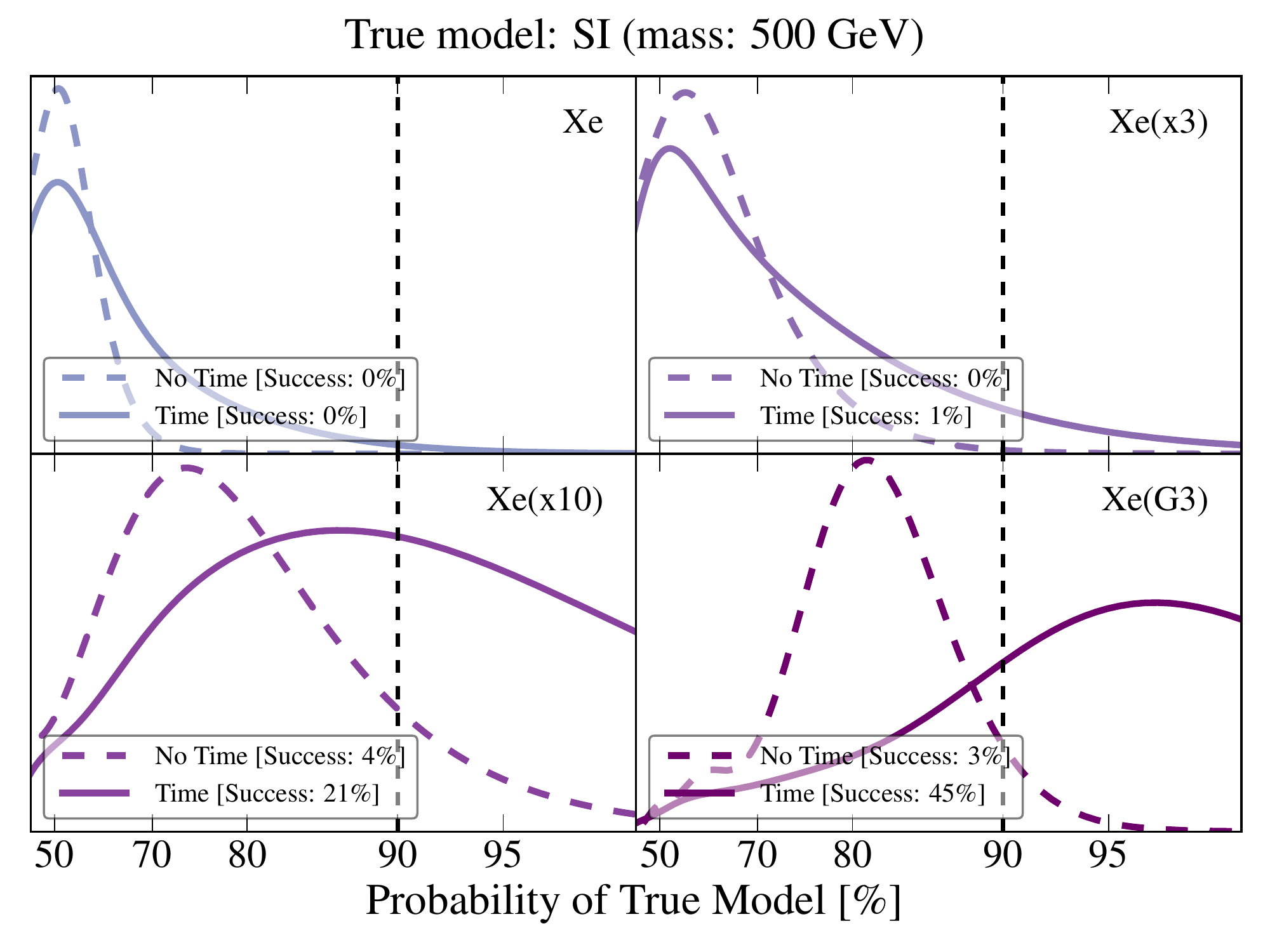}
\caption{\label{fig:500gev_SI_Higgs_XeFull_TNT_GF}
Same as Fig.~\ref{fig:20gev_anapole_XeFull_TNT_GF} but for a $500$ GeV DM and the SI interaction as the true model.}
\end{figure*}

We present in Figs.~\ref{fig:20gev_SI_Higgs_XeFull_TNT_GF}--\ref{fig:500gev_SI_Higgs_XeFull_TNT_GF} the model selection prospects for various exposures on a xenon experiment, including (solid lines) and neglecting (dashed lines) information on the modulation of the recoil rate, and assuming the SI interaction is the true model. Results are shown for $20$ GeV (\Fig{fig:20gev_SI_Higgs_XeFull_TNT_GF}), $125$ GeV (\Fig{fig:125gev_SI_Higgs_XeFull_TNT_GF}), and $500$ GeV (\Fig{fig:500gev_SI_Higgs_XeFull_TNT_GF}) DM particle. Results are similar to those presented in \Sec{sec:results} for the case of the anapole interaction.

\bibliographystyle{JHEP}
\bibliography{mod-sel}

\end{document}